\documentclass[useAMS]{mn2e}
\usepackage{graphicx}
\usepackage{longtable}

\title[NGC 6822]{The Local Group Galaxy NGC 6822 and its Asymptotic Giant
Branch Stars}

\author[Whitelock et al.]{Patricia A. Whitelock$^{2,1}$, John W.
Menzies$^{2}$, Michael W. Feast$^{1,2}$, \newauthor Francois
Nsengiyumva$^{2,1}$, and Noriyuki Matsunaga$^3$\\
      $^1$ Astronomy, Cosmology and Gravity Centre, Astronomy Department,
           University of Cape Town, 7701 Rondebosch, South Africa.\\
      $^2$ South African Astronomical Observatory, P.O.Box 9, 7935
           Observatory, South Africa.\\
      $^3$ Kiso Observatory, Institute of Astronomy, The University of Tokyo,
           10762-30, Mitake, Kiso, Nagano 397-0101, Japan.\\
 }

\begin{document}
\maketitle
\begin{abstract}
 $JHK_{S}$ photometry is presented from a 3.5 year survey of the central
regions of the irregular galaxy NGC\,6822.  The morphology of the
colour-magnitude and colour-colour diagrams is discussed with particular
reference to M, S and C-type AGB stars and to M-supergiants.  Mean $JHK_{S}$
magnitudes and periods are given for 11 O-rich and 50 presumed C-rich Miras. 
Data are also listed for 27 large amplitude AGB stars without periods and
for 69 small amplitude AGB variables.  The slope of the bolometric
period-luminosity relation for the C-rich Miras is in good agreement with
that in the LMC.  Distance moduli derived from the C- and O-rich Miras are
in agreement with other estimates.  The period distribution of C-rich Miras
in NGC\,6822 is similar to that in the Magellanic Clouds, but differs from
that in the dwarf spheroidals in the Local Group.  In the latter there is a
significant proportion of large amplitude, short period variables indicating
a population producing old carbon-rich AGB stars.
 \end{abstract}
\section{Introduction}

NGC\,6822 is an isolated barred dwarf irregular galaxy within the Local
Group.  It is comparable in size to the SMC but has a slightly higher
metallicity (Muschielok et al. 1999; Venn et al.  2001). It contains
numerous supergiants and H{\sc ii} regions with obvious signs of star
formation and is sometimes referred to as a polar ring galaxy (Demers et al.
2006).  Recent HST colour-magnitude studies suggest that over 50 percent
of its stars formed in the last 5\,Gyr (Cannon et al. 2012). 

We present here new multi-epoch $JHK_{S}$ photometry of the central regions
of NGC\,6822 which we compare with earlier work and use to identify and
characterize large amplitude, Mira, asymptotic giant branch (AGB) variables. 
Earlier papers used the same photometry to identify the first symbiotic star
in NGC\,6822 (Kniazev et al.  2009) and to study the Cepheid variables
(Feast et al.  2012). A preliminary analysis of the AGB variables was
produced by Nsengiyumva (2010).

Earlier studies of red giants and AGB stars in NGC\,6822 have been made by
Cioni \& Habing (2005), Kang (2006), Groenewegen et al.  (2009), Kacharov,
Rejkuba \& Cioni (2012) and by Sibbons et al.  (2012), while Battinelli \&
Demers (2011) specifically identified AGB variables.

This work forms part of a broad study of AGB variables in Local Group Galaxies
which so far has covered Leo~I (Menzies et al.  2002; 2010), Phoenix
(Menzies et al.  2008), Fornax (Whitelock et al.  2009) and Sculptor
(Menzies et al.  2011).   These new observations provide an opportunity to
compare these dwarf spheroidals with a dwarf irregular surveyed in the same
way. 

\section{Observations} 
 Our survey of NGC\,6822 is confined to the optical bar which is aligned
nearly N-S.  We used the Japanese-South African IRSF telescope equipped with
the SIRIUS camera, which permits simultaneous imaging in the $J, H$ and
$K_S$ bands (see Nagayama et al.  (2003) for details).  We defined 3
overlapping fields, with field 1 centred at $\alpha$(2000.0) =
$19^h44^m56^s$ and $\delta$(2000.0) =$-14^o48'06''$.  Fields 2 and 3 are
centred 6.7 arcmin N and S, respectively, of field 1.  The three fields,
each approximately 7.8 arcmin square, were observed in $JHK_S$ at 19, 18 and
16 epochs, respectively, over a period of 3.5 years.
%Typically 30 dithered exposures of 30 s each were combined at each epoch,
%though occasionally, depending on sky brightness at $K_S$, exposure times
%were reduced to 20 s.

\begin{table*}
%\tiny
\caption[]{Data for stars with standard errors less than 0.1 mag (note that
this selection omits the large amplitude variables which are listed in
various tables below). The full table
is available on-line. The first two columns are the
equatorial coordinates in degrees; N is our own
identification number; the mean photometry, $JHK_S$ is listed together with
its standard deviation, $\delta JHK_S$; NJ, NH and NK are the number of
observations used to derive the means.}
\begin{center}
\begin{tabular}{ccccccccccccccc}
\hline
RA & Dec & N & $J$& $\delta J$& $H$ & $\delta  H$& $K_S$ & $\delta  K$ &
$J-H$ & $H-K_S$ & $J-K_S$ &  NJ & NH & NK\\
\multicolumn{2}{c}{(2000.0)}&& \multicolumn{9}{c}{(mag)}\\
\hline
296.28656 &--14.80424 &10001  &12.613 &  0.009 & 12.050 &  0.006 & 11.939  & 0.026 &  0.563 &  0.111 &  0.674 & 18 &14 &18\\
296.17661 &--14.78323 &10002  &12.704 &  0.014 & 12.303 &  0.016 & 12.217  & 0.030 &  0.401 &  0.086 &  0.487 & 18 &18 &18\\
296.18298 &--14.83667 &10008  &13.354 &  0.018 & 12.903 &  0.008 & 12.819  & 0.033 &  0.451 &  0.084 &  0.535 & 18 &16 &18\\ 
296.20160 &--14.83465 &10009  &13.602 &  0.009 & 13.112 &  0.004 & 13.020  & 0.021 &  0.490 &  0.092 &  0.582 & 18 &12 &18\\
296.21399 &--14.82684 &10010  &13.161 &  0.009 & 12.751 &  0.009 & 12.673  & 0.012 &  0.410 &  0.078 &  0.488 & 17 &17 &18\\
\hline
\end{tabular}
\end{center}
\label{tab_main} 
\end{table*}

Further details, including those of the photometric calibration, are given
by Feast et al.  (2012).  The basic data for stars with 
standard errors less than 0.1 mag in each band are provided on-line, and the
first few lines of the catalogue are illustrated in 
Table~\ref{tab_main} (The Mira variables, discussed in section 5, are not in
this table).    Numbers of observations, NJ, NH etc.  larger than 19
are found for some stars in the areas of overlap between fields.

NGC\,6822 is at low galactic latitude, $b=-18^{\circ}.4$, so it experiences
some interstellar extinction as well as confusion with Galactic sources. 
For the interstellar extinction we adopt $A_V=0.77$ mag (amounting to
$A_J=0.20$, $A_H=0.12$, $A_K=0.07$ mag) from Clementini et al.  (2003) using the
information from Schlegel, Finkbeiner \& Davis (1998).  We note, however,
that the extinction across NGC\,6822 is somewhat variable and that
significantly higher values are possible for sources associated with star
forming regions.  Our discussion of the AGB, and of the large amplitude
variables in particular, will not be very sensitive to either the reddening,
or its variability.

\section{Colour-Magnitude Diagram}

\begin{figure*}
\includegraphics[width=17.6cm]{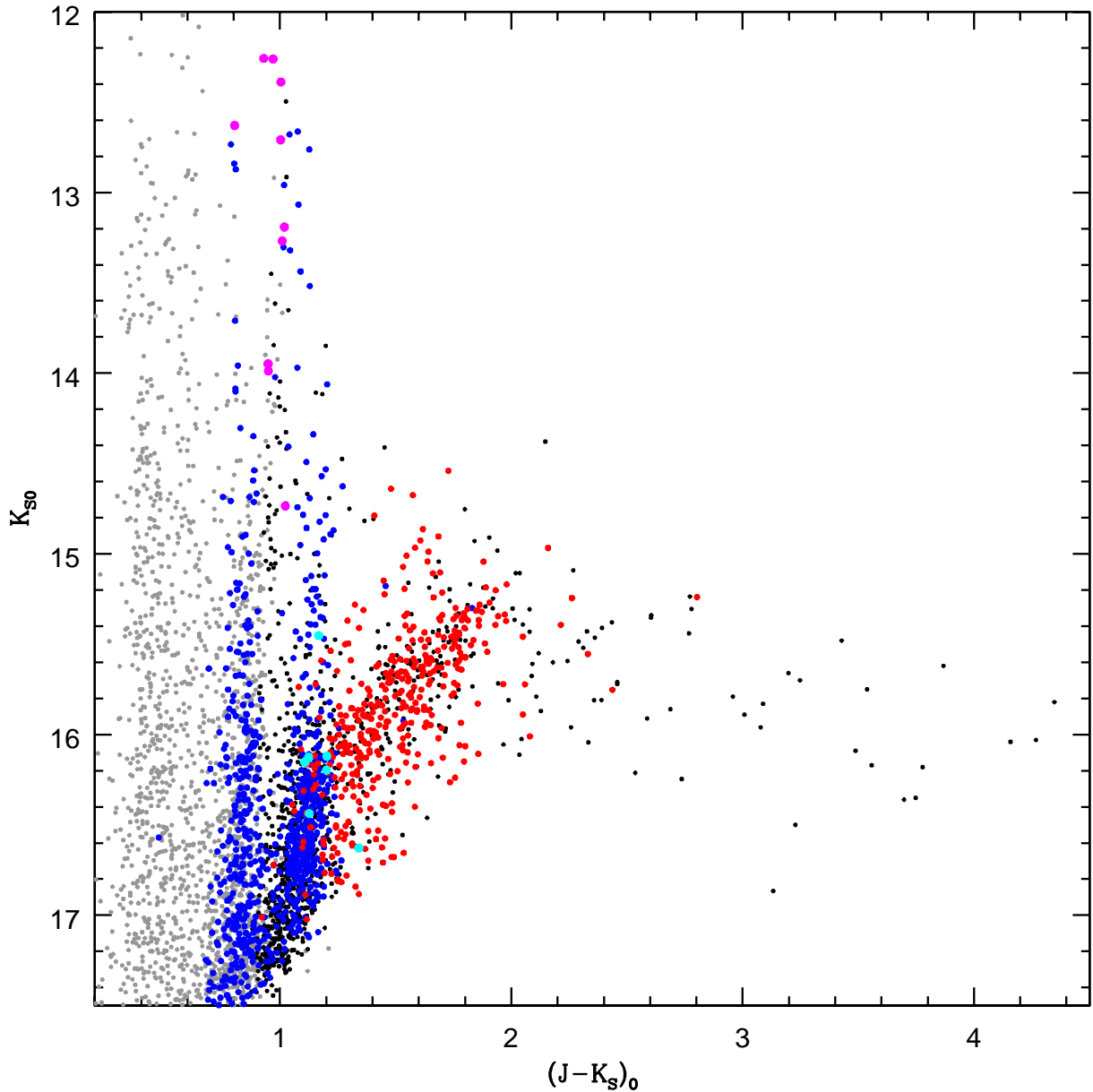}
\caption{Colour-magnitude diagram showing the stars in our NGC\,6822
catalogue in black and, for those that are probably foreground dwarfs, in
gray (unless they have spectral types). The TRGB is at $K_0=17.42$ mag. 
Stars of known late spectral-type are
shown in colour: those with narrow-band colours of M stars  
and C stars are shown in blue and red, respectively; S stars are shown
in cyan and M supergiants in magenta.}
\label{fig_cm1}
\end{figure*}

\begin{figure*}
\includegraphics[width=17.6cm]{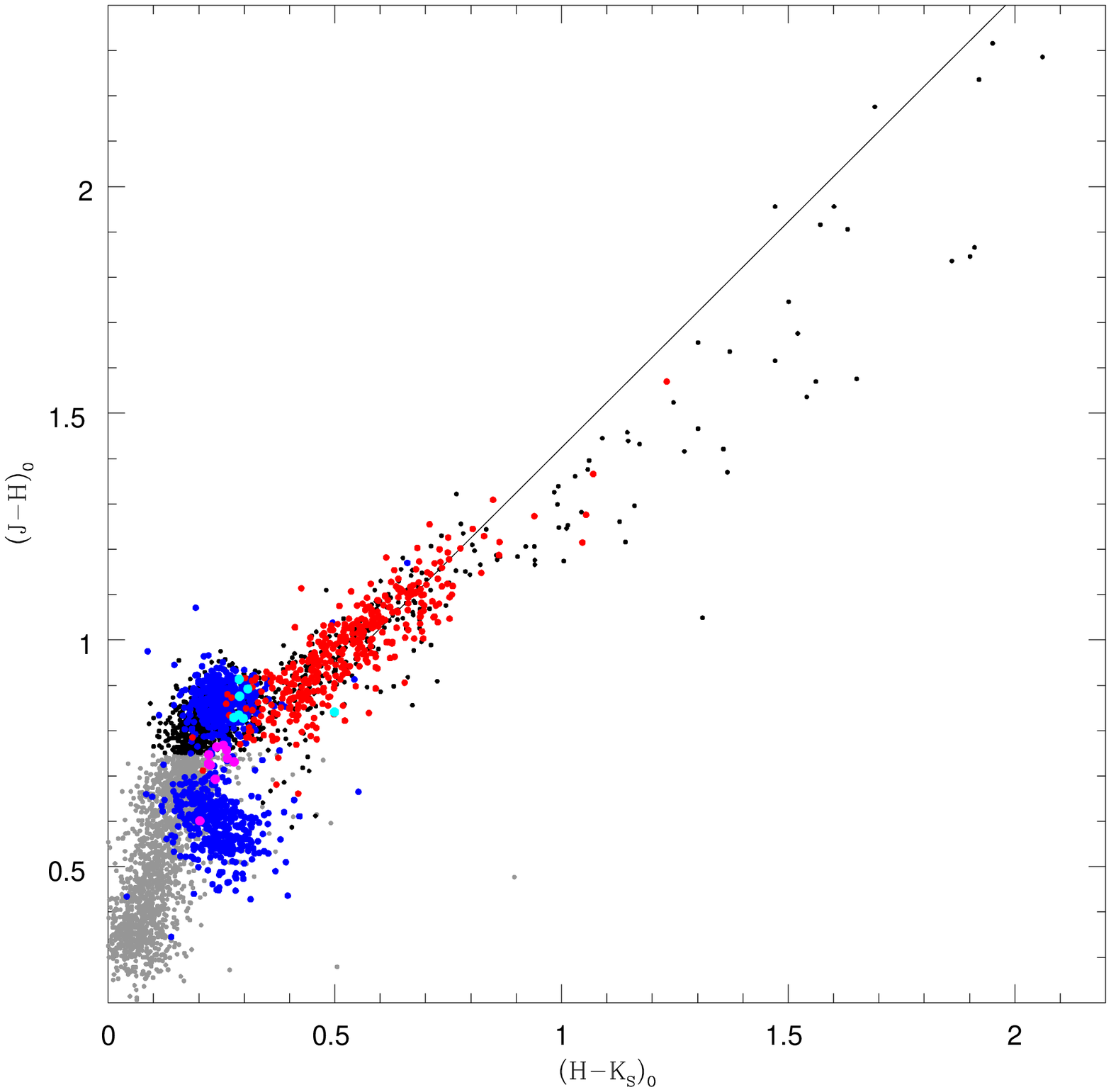}
 \caption{Two-colour diagram with the same stars as in Fig.~\ref{fig_cm1}. 
The line represents the locus for Galactic carbon Miras (equation 2 from
Whitelock et al.  (2006) converted to 2MASS as: $(H-K_S)_0=
1.003(J-H)_0-0.428$.  Note the very clear separation in $(J-H)_0$ of the
foreground M-dwarfs (smaller values) and NGC\,6822 M-giants (larger values). 
The M supergiants have intermediate colours.  } 
\label{fig_cc1} 
\end{figure*}

The colour-magnitude and two-colour diagrams for stars from
Table~\ref{tab_main} plus the variables discussed in section 5, are illustrated in
Fig.~\ref{fig_cm1} and Fig.~\ref{fig_cc1}.

According to the detailed analysis by Sibbons et al. (2012) the tip of the
red giant branch (TRGB) is at $K_0=17.42\pm0.11$ mag (2MASS system).  So
here we are dealing entirely with AGB stars, together with a sprinkling of
red supergiants at the highest luminosity.  Following Sibbons et al.  we
identify stars with $(J-H)_0<0.75$ mag as most likely to be foreground dwarfs
and they are
% therefore 
shown as gray, unless they have spectral types. 
Most, but not all, of the stars populating the left of the colour magnitude
diagram and the bottom of the two-colour diagram are foreground dwarfs.

The morphology of these diagrams is more clearly understood when stars of
known spectral type are identified and it is therefore discussed in the next
section.


\section{Spectral Types}
 Various groups have attempted to separate the AGB stars into M- and C-type
on the basis of their $J-K$ colour (Cioni \& Habing (2005), Kang et al.
(2006), Sibbons et al. (2012)). Kacharov et al. (2012) obtained spectra,
including 148 stars in common with us, two of which are Mira variables. 
They concluded that 79 percent of the carbon stars had $(J-K)_0>1.28$ mag
(on the 2MASS system). 

Letarte et al. (2002) obtained narrow band photometry over a very large
field in order to find the extent of the NGC\,6822 C-star population.  
%We searched their catalogue for stars in common with our sample, using a 1
%arcsec search radius, after adding a correction of $-0.38$ arcsec to their
%declinations (which experiment indicated was the mean difference between the
%two coordinate systems).
Over 5000 stars were found in common with our sample including
about 1000 with the colours of M stars ($R-I>1.1$ and $CN-TiO<0$) and 430
with the colours of C stars ($R-I>1.1$ and $CN-TiO>0.3$) and this allows a
reasonably good division between M- and C-type stars among our variables
(section 5).  However, the central field is crowded and a few
misidentifications are possible and might explain some outlying points among
the C- and M- stars.

 Levesque \& Massey (2012) discuss red supergiants (RSGs) in NGC\,6822 and
use the $V-R$ and $B-V$ colours to separate RSGs from giants. 
Table~\ref{tab_super} shows the photometry of the M-type supergiants from
Levesque \& Massey's table~2, which are also illustrated in
Figs.~\ref{fig_cm1} and \ref{fig_cc1}.  N10032 is a small amplitude variable
(as the uncertainty on the mean magnitudes listed in Table~\ref{tab_super}
show).  It is not obviously periodic and the full peak-to-peak amplitude is
$\Delta K_S \sim 0.3$ mag.  Note also that N10015 falls amongst the
dwarfs in both figures, although its status as a supergiant is well
established.  As others have noted (Cioni \& Habing 2005, see also Nikolaev
\& Weinberg 2000) there is no clear distinction in the
colour-magnitude diagram between the giants and supergiants (or even between
the supergiants and foreground dwarfs).

The broad morphology of Fig.~\ref{fig_cm1} is now clear. The vertical strip
between $(J-K)_0\sim 0.3$ and 0.6 mag is mostly foreground stars, but will
include warm supergiant members of NGC\,6822, e.g., the Cepheids (section
5).  The vertical strip around $(J-K_S)_0\sim 0.9$ mag is mostly foreground
M-dwarfs.  The almost vertical strip at around $(J-K)_0\sim 1.1$ mag starts
just above the TRGB as M stars on the AGB; at higher luminosity,
$K_{S0}<15.8$ mag, it becomes luminous AGB stars.  These stars are discussed
further in the context of the variable stars (section 5), but are presumably
younger than the bulk of the AGB population that evolve to the right of the
diagram as C stars.  It is here we would expect to find hot bottom burning
stars (e.g.  Sackman \& Boothroyd 1992) and super-AGB stars (e.g.  Siess
2008), prior to the onset of heavy mass-loss.  At even higher luminosities,
and between the AGB column and the M-dwarf column, fall the M supergiants. 
The carbon stars concentrate in a diagonal band to the right of the AGB
M-type stars.  The AGB variables without spectral types\footnote{most of
these will be C stars, but this is also the place we expect to find OH/IR
stars if such objects exist in NGC\,6822.} extend the C-star sequence to the
extreme right of the diagram.  It is likely that a small number of the
points below the carbon stars ($K_S>16$ mag) in Fig.~\ref{fig_cm1} are actually
unresolved galaxies (see e.g.  Whitelock et al.  2009).

\begin{table*}
\caption[]{M Supergiants from Levesque \& Massey (2012).}
\begin{center}
\begin{tabular}{ccccccccccccccl}
\hline
LGGS& RA& Dec & N & $J$& $\delta J$& $H$ & $\delta  H$& $K_S$ & $\delta  K$ 
& $J-K_S$ & NJ & NH& NK & Sp\\
& \multicolumn{2}{c}{(2000.0)}&& \multicolumn{7}{c}{(mag)}\\
\hline
J194445.76-145221.2 &  296.19067& -14.87276& 30016& 13.91& 0.01& 13.09& 0.03& 12.78& 0.02& 1.13& 12& 14& 11   &M1\\ 
J194447.81-145052.5 &  296.19919& -14.84817& 40115& 14.41& 0.03& 13.57& 0.03& 13.26& 0.02& 1.15& 24& 26& 22    & M1\\
J194450.44-144410.0 &  296.21021& -14.73628& 40177& 15.14& 0.01& 14.33& 0.03& 14.06& 0.02& 1.08& 22& 24& 22    & M2\\
J194453.46-144540.1 &  296.22278& -14.76476& 10089& 15.10& 0.03& 14.29& 0.02& 14.02& 0.02& 1.08& 18& 17& 17    & M4.5\\
J194454.46-144806.2 &  296.22696& -14.80191& 10032& 14.48& 0.05& 13.66& 0.03& 13.34& 0.05& 1.14& 14& 14& 16   &M1\\
J194454.54-145127.1 &  296.22726& -14.85778& 40278& 13.43& 0.03& 12.60& 0.04& 12.33& 0.05& 1.10& 33& 33& 33   &M0\\
J194455.70-145155.4 &  296.23212& -14.86564& 40315& 13.39& 0.06& 12.61& 0.05& 12.33& 0.05& 1.06& 27& 26& 28   &M0\\
J194457.31-144920.2 &  296.23883& -14.82247& 10011& 13.60& 0.04& 12.75& 0.04& 12.46& 0.05& 1.14& 18& 18& 18   &M1\\
J194459.86-144515.4 &  296.24945& -14.75443& 10015& 13.64& 0.00& 12.95& 0.01& 12.70& 0.02& 0.94& 13& 17& 18   &M1\\
J194503.58-144337.6 &  296.26492& -14.72723& 20101& 15.96& 0.02& 15.11& 0.02& 14.81& 0.02& 1.15& 16& 17& 16   &M0\\
\hline
\end{tabular}
\end{center}
\label{tab_super} 
\end{table*}

\begin{table*}
\caption[]{Stars with known S spectral-type.}
\begin{center}
\begin{tabular}{ccccccccccccc}
\hline
RA & Dec & N & $J$& $\delta J$& $H$ & $\delta  H$& $K_S$ & $\delta  K$  
&  $J-K_S$ & NJ & NH& NK \\
 \multicolumn{2}{c}{(2000.0)}&& \multicolumn{7}{c}{(mag)}\\
\hline
296.17892 & --14.82286 & 10870 & 17.52& 0.04& 16.53& 0.04& 16.19& 0.05& 1.34& 15& 17& 17\\
296.21545 & --14.83469 & 10784 & 17.45& 0.03& 16.53& 0.04& 16.20& 0.06& 1.26& 17& 18& 17\\
296.27341 & --14.80861 & 11004 & 17.60& 0.03& 16.62& 0.03& 16.27& 0.05& 1.33& 15& 16& 15\\
296.28308 & --14.80497 & 11029 & 17.46& 0.03& 16.55& 0.02& 16.22& 0.05& 1.24& 16& 16& 17\\
296.25427 & --14.81764 & 12050 & 18.17& 0.07& 17.25& 0.10& 16.70& 0.05& 1.47& 14& 17& 17\\
296.19156 & --14.89296 & 30528 & 17.76& 0.06& 16.85& 0.04& 16.51& 0.10& 1.26& 12& 12& 14\\
296.25522 & --14.82579 & 10326 & 16.82& 0.03& 15.86& 0.03& 15.52& 0.02& 1.30& 16& 16& 16\\
\hline
\end{tabular}
\end{center}
\label{tab_s} 
\end{table*}

\subsection {S stars}
 Six of the nine S stars identified by Kacharov et al. (2012) fall in the
area we surveyed and they are listed in Table~\ref{tab_s}.  All of these
have the $K_S$ magnitudes, and all but one have the colours, anticipated for
an evolutionary state between that of the lower luminosity M stars and the
C stars (see Fig.~\ref{fig_cm1} and \ref{fig_cc1}).  The exception, N12050,
has a slightly redder $J-K_S$ and therefore falls amongst the C stars (as
noted by Kacharov et al.).  The S star identified by Aaronson et al.  (1984)
is our N10326 which is about a magnitude brighter than the Kacharov et al. 
S stars.  If it is an intrinsic S star (i.e.  its s-process elements are the
consequence of its own evolution and dredge-up and are not from a close
companion) then it must be more massive than the others, perhaps comparable
to the hot bottom burning Li-rich S stars in the LMC and SMC (Smith et al. 
1995; Whitelock et al.  2003).  In that case it may be from the same
population as the luminous large-amplitude O-rich variables discussed below.

\begin{figure*}
\includegraphics[width=17.6cm]{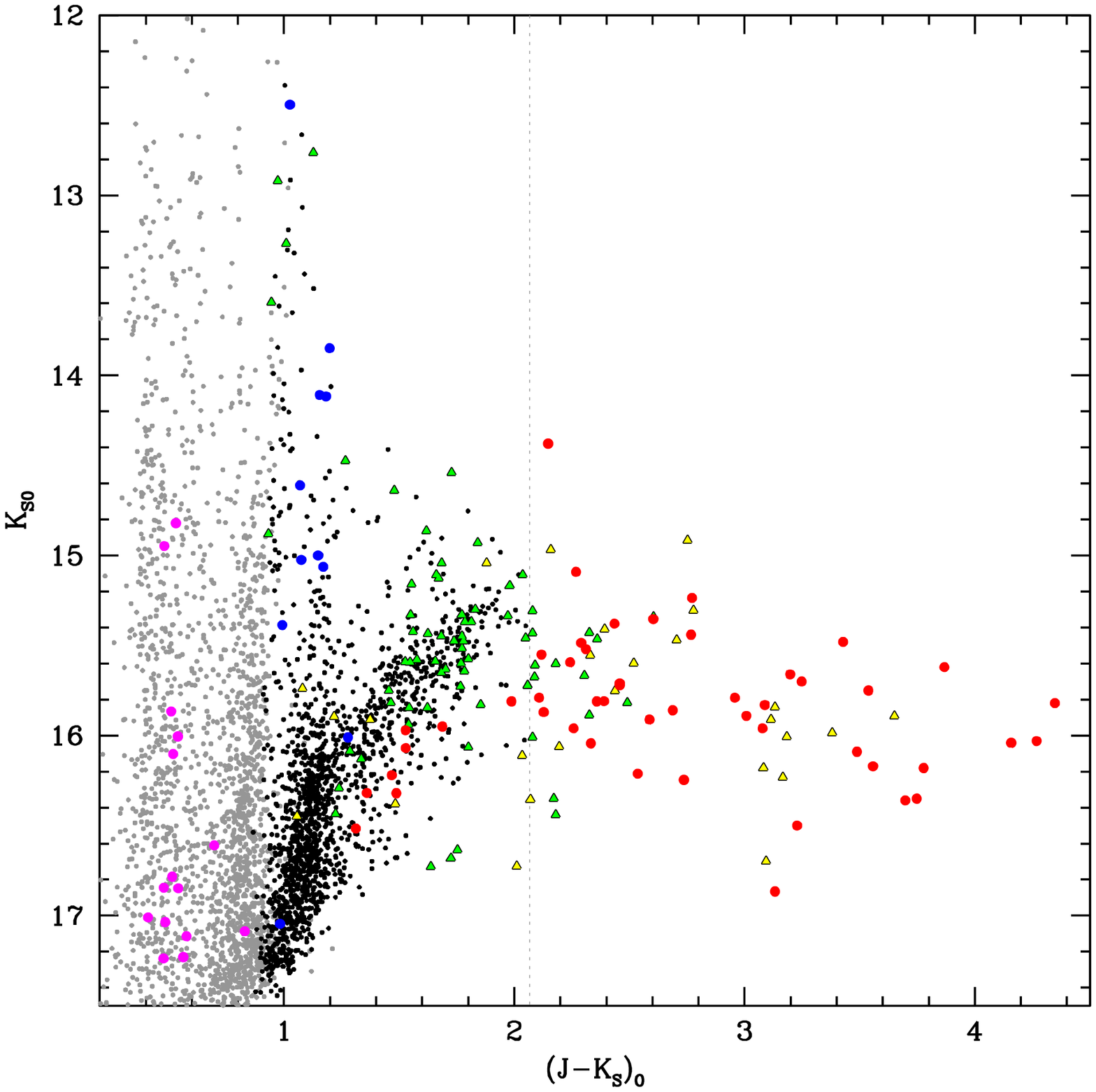}
\caption{Colour-magnitude diagram illustrating the variable stars.
Symbols: Cepheids magenta; M-type Miras and supergiant blue; C-type Miras red; 
other variables triangles: large amplitude yellow, small amplitude green. All
stars to the right of the dotted line were systematically examined for variability.}
\label{fig_cm2}
\end{figure*}
\begin{figure*}
\includegraphics[width=17.6cm]{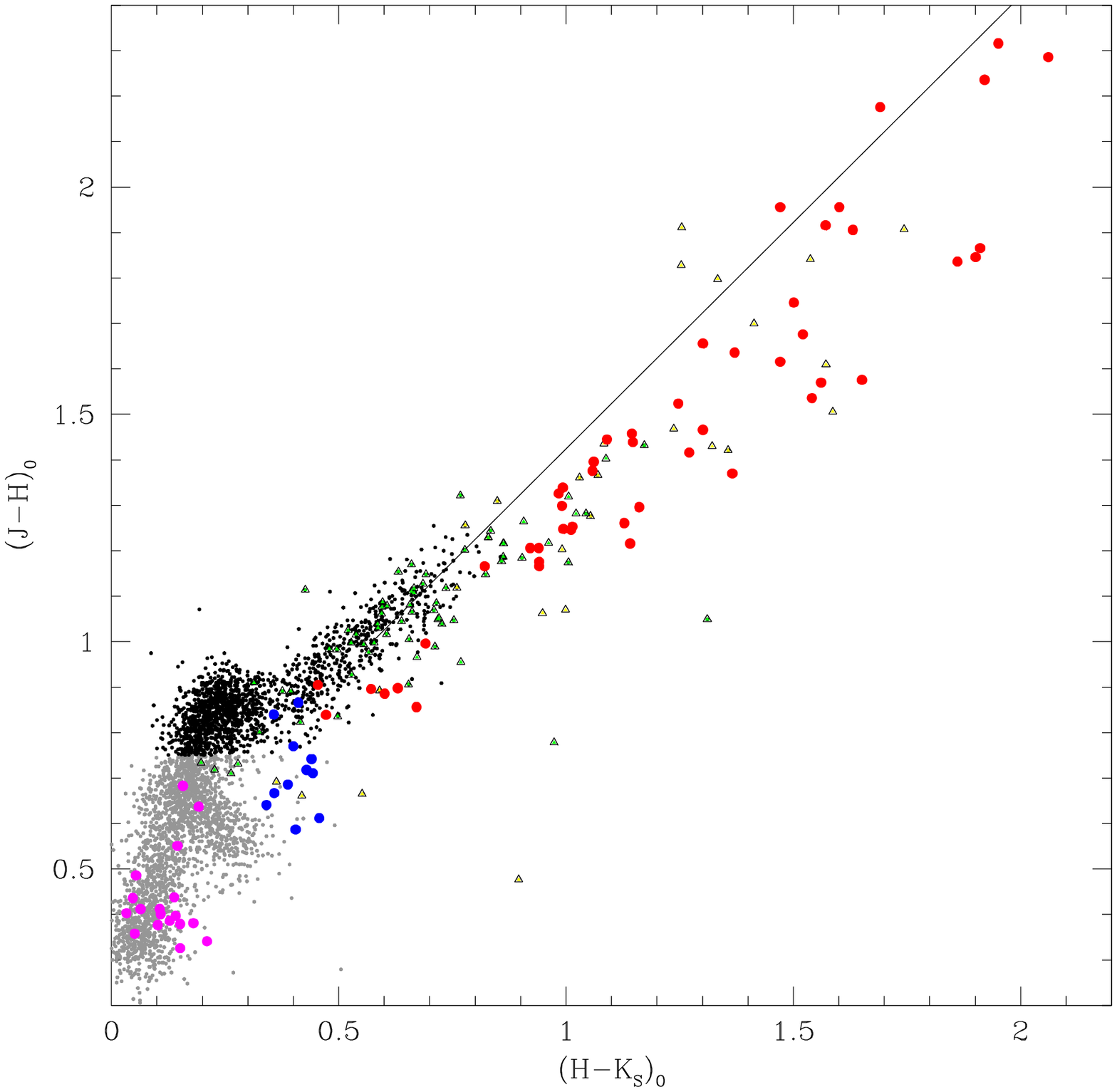}
\caption{Two-colour  diagram showing the same stars as in
Fig.~\ref{fig_cm2}.}
\label{fig_cc2}
\end{figure*}

\section{Variables}
 We examined the light curves of all stars with $K_S<17$ mag for which we had at
least 10 observations and which showed a standard deviation in $J$, $H$ or
$K_S$ of $>0.2$ mag, going to lower standard deviations for brighter
magnitudes.  By this approach we found the brightest of the Cepheids, all of
the stars listed in Tables~\ref{tab_o}(a) and (b) plus a considerable
fraction of those which we list in the later tables and which are discussed
below.  Given that our primary objective was to find large amplitude AGB
variables we then examined stars with $J-K_S>2.2$ mag, finding them all to be
variable at some level.

 It should be noted
that one consequence of the use of a reference frame in the $H$ band to
provide positions at which ``fixed-position" DoPHOT photometry was performed
(Feast et al.  2012) is that extremely red stars or stars with a very large
amplitude of variation might have been missed.  If the star was not
measurable on the reference $H$ frame, then it would not have been found in
any other band either.  This means that there may be red variables that we
have not measured.  The limiting magnitudes at $J, H$ and $K_S$ in our
catalogue are approximately 20.3, 18.3 and 18.0 mag, respectively.  This
means that we would have missed red variables with $H-K_S=2.0$ mag that were
fainter than about $K_S=16.3$ mag or $J=18.4$ mag at the time the reference frame
was obtained, even though these latter values are significantly above the
relevant limiting magnitudes.

The various variables are identified in Figs.~\ref{fig_cm2} and
\ref{fig_cc2}. The Cepheids were discussed by Feast et al.  (2012) and the
others are considered below.

Periods were determined by Fourier analysis and Table~\ref{tab_o} lists the
Fourier mean $JHK_S$ magnitudes for the large amplitude (see below)
variables with measurable periods (Miras), together with the peak-to-peak
amplitudes.  The table is split into (a) O-rich, M-type, stars and (b)
C-stars, on the basis of $J-K_S$ colour (see section 4).  Table~\ref{tab_o}
is the only one to list Fourier mean magnitudes.  The other tables contain
simple mean values of all the observations.  Note that the stars in
Table~\ref{tab_o} are \underline{not} in the online catalogue, because their
mean magnitudes are evaluated differently.

Our previous practice has been to define large amplitude, Mira, variables to
be those with $\Delta K_S>0.4$ mag (e.g.  Whitelock et al.  2006).  While
the distinction between Miras and SR variables is clear for O-rich stars
(Miras were originally defined from observations of Galactic stars, most of
which are O-rich), it is not so clear for the C-rich stars (Whitelock 1996)
and it is apparent that this cut-off results in far fewer short period
($P<300$ days) Miras than we might expect in NGC\,6822, and presumably in
similar galaxies.  However, if we are interested in these variables as
distance indicators the distinction is important, because low-amplitude
variables can fall on any one of several period-luminosity (PL) relations
(Wood 2000, Ita et al.  2004) whereas the Miras with periods less than
450\,days fall only on one relation.  Nevertheless, we relax the
criterion very slightly here to include stars with $0.36<\Delta K_S<0.4$ mag,
while noting that we have done so in Table~\ref{tab_c}(b).  The $K_S$ light
curves of the Mira variables are illustrated in the appendix.

With regard to establishing the O- or C-rich nature of the Miras, we
follow Letarte at al. (2002) and relaxing their criteria (as they suggest)
very slightly to examine stars with $R-I>1.0$ (rather than 1.1) we find the
following: 6 Miras have M-type narrow-band colours; they are labeled `M' in the
last column of Table~\ref{tab_o}(a), and all have $(J-K_S)_0<1.33$ mag. 
Similarly, 11 stars have C-type narrow-band colours; these are labeled `C'
in the last column of Table~\ref{tab_c}(b), and all have $(J-K_S)_0>1.33$
mag.

Two of the Miras have spectral types in Kacharov et al. (2012), as indicated
in Table~\ref{tab_c}(b).  N\,12790 (P=182 days) has spectral type C6.5 and is the
bluest ($(J-K_S)_0=1.31$ mag) of the stars that we consider to be C stars in our
PL relation analysis (see section 7 below).

Table \ref{tab_var1} contains large amplitude (mostly $\Delta K_S>0.4$ mag)
but not obviously periodic variables; it includes some Miras for which we
could not estimate periods and probably some unrecognized Miras (see below). 
Table~\ref{tab_var2} contains small amplitude ($\Delta K_S<0.4$ mag)
variables, most of which are not unambiguously periodic; these are probably
semi-regular (SR) or irregular variables.  This group includes a few which
have SR-variations superimposed on an apparently secular trend, which can
make the overall change more than 0.4 mag.  These are candidates for
variables with two periods.

The distinction between Tables \ref{tab_var1}  and \ref{tab_var2}  is to
some extent subjective and there is undoubtably overlap between the two
groups. Nevertheless, they do show distinctly different colours (see
Figs.~\ref{fig_cm2} and \ref{fig_cc2}), with most of the large amplitude
variables having larger values of $(J-K_S)$ than those with small
amplitudes, suggesting that they have higher mass-loss rates.  

Given the cadence of our observations and the fact that at least some
mass-losing C-rich Miras undergo periods of very erratic variation (e.g. 
R~For as illustrated in Whitelock et al.  (1997) and several LMC C-rich
Miras discussed by Whitelock et al. (2003)) we anticipate that
a significant fraction of the stars in Table~\ref{tab_var1} will be
Miras of this type (see also N21029 in Fig.~A2 in the appendix). 

There are a few of the low amplitude variables among the supergiants.  Note
also that many of the AGB stars and supergiants not identified as variables
will in fact be low amplitude variables, generally with $\Delta K_S < 0.2$ mag.

\begin{table*}
\caption[]{(a) Periodic Large Amplitude O-Rich Variables (Miras).
Fourier mean magnitudes $\bar{J}$, $\bar{H}$, $\bar{K}$ are listed with
the peak-to-peak amplitudes ($\Delta J$, $\Delta H$, $\Delta K_S$) 
of the best fitting first-order sine curves.  The last column shows those
with spectral types, or narrow band spectral indices (Letarte et al.  2002).}
\begin{center}
\begin{tabular}{ccccccccccc}
\hline
RA & Dec & N & $P$ & $\bar{J}$ & $\bar{H}$ & $\bar{K_S}$ & $\Delta  J$ & $\Delta  H$ & $\Delta  K_S$
& Sp\\
\multicolumn{2}{c}{(2000.0)}&& (days)& \multicolumn{6}{c}{(mag)}\\
\hline
 296.18398 & --14.78018 & 12557&  158 & 18.23 & 17.51 & 17.12 & 0.84 & 0.83 & 0.80& M\\
 296.25229 & --14.78475 & 11226&  257 & 17.49 & 16.54 & 16.08 & 0.54 & 0.58 & 0.53&M:\\
 296.20415 & --14.63486 & 20331&  314 & 16.58 & 15.91 & 15.46 & 0.88 & 1.10 & 0.86& M \\
 296.22364 & --14.77473 & 10184&  370 & 16.44 & 15.58 & 15.13 & 0.81 & 0.95 & 0.89& M \\
 296.21816 & --14.88035 & 30133&  401 & 16.30 & 15.53 & 15.09 & 0.85 & 1.02 & 0.94 \\
 296.21322 & --14.68097 & 20134&  402 & 16.35 & 15.55 & 15.07 & 0.89 & 1.00 & 0.92 \\
 296.20428 & --14.74271 & 40139&  545 & 15.25 & 14.32 & 13.92 & 0.62 & 0.72 & 0.66 \\
 296.25088 & --14.76786 & 10198&  602 & 15.50 & 14.68 & 14.19 & 0.75 & 0.93 & 0.80 \\
 296.18801 & --14.87231 & 30292&  637 & 15.88 & 15.19 & 14.68 & 0.96 & 1.00 & 0.97& M \\
 296.26702 & --14.76311 & 10091&  638 & 15.46 & 14.67 & 14.18 & 0.95 & 1.12 & 1.00 &(1)\\
 296.21293 & --14.73224 & 20004&  854 & 13.72 & 12.97 & 12.57 & 0.47 & 0.45 & 0.40& M \\

\hline
\end{tabular}
\end{center}
(1) N10091 was identified as a late-M star with $ \rm H\alpha$ emission by Filippenko
\& Chornock (2003).
\label{tab_o} 
\end{table*}

\setcounter{table}{3}
\begin{table*}
\caption[]{(b) Periodic Large Amplitude C-Rich Variables (Miras). Columns as
given for Table~\ref{tab_o}a.}
\begin{center}
\begin{tabular}{cccclcccccl}
\hline
N & RA & Dec &$P$ & $\bar{J}$ & $\bar{H}$ & $\bar{K_S}$ & $\Delta  J$ & $\Delta  H$ & $\Delta  K_S$
& Sp\\
& \multicolumn{2}{c}{(2000.0)}& (days)& \multicolumn{6}{c}{(mag)}\\
\hline
12790 &  296.20389 & -14.75535 & 182 & 18.03 & 17.11 & 16.59 & 1.09 & 0.82 & 0.51 & C6.5(1)\\
10817 &  296.23639 & -14.83054 & 214 & 17.70 & 16.76 & 16.04 & 0.75 & 0.61 & 0.45 &  \\
20540 &  296.18011 & -14.71067 & 223 & 17.88 & 16.89 & 16.39 & 0.94 & 0.71 & 0.48 & C\\
40590 &  296.28567 & -14.73952 & 223 & 18.01 & 17.04 & 16.39 & 0.80 & 0.63 & 0.39 & C *\\
12751 &  296.21027 & -14.75973 & 231 & 17.89 & 16.91 & 16.29 & 1.07 & 0.72 & 0.50 & \\
11032 &  296.18048 & -14.80431 & 239 & 18.20 & 16.91 & 15.94 & 1.20 & 0.92 & 0.65 &  \\
10748 &  296.17438 & -14.83893 & 243 & 18.58 & 17.16 & 16.11 & 1.06 & 0.77 & 0.48 &  \\
20578 &  296.29065 & -14.69641 & 246 & 17.80 & 16.82 & 16.14 & 0.77 & 0.79 & 0.43 & C\\
20542 &  296.17703 & -14.71031 & 255 & 17.84 & 16.76 & 16.02 & 0.73 & 0.66 & 0.50 & C\\
30430 &  296.21802 & -14.92573 & 269 & 17.67 & 16.56 & 15.86 & 0.87 & 0.63 & 0.38 & C *\\
12208 &  296.23352 & -14.80653 & 278 & 17.87 & 16.61 & 15.62 & 1.34 & 1.07 & 0.76 &  \\
21419 &  296.27563 & -14.74923 & 278 & 19.93 & 18.27 & 16.57 & 1.50 & 1.42 & 1.05 &  \\
13364 &  296.19489 & -14.82267 & 286 & 18.10 & 16.85 & 15.86 & 1.30 & 1.09 & 0.75 &  \\
12400 &  296.17468 & -14.79389 & 301 & 18.00 & 16.75 & 15.88 & 1.11 & 0.84 & 0.60 & C\\
30583 &  296.30014 & -14.87872 & 302 & 18.37 & 17.07 & 15.88 & 1.2 & 1.1 & 0.5 & C (4)\\ 
20239 &  296.22632 & -14.72554 & 304 & 18.04 & 16.71 & 15.66 & 0.86 & 0.67 & 0.48 &  \\
20558 &  296.19647 & -14.70265 & 304 & 18.21 & 16.60 & 15.31 & 0.41 & 0.40 & 0.44 & (2)\\
20840 &  296.30347 & -14.74166 & 306 & 18.7 & 17.18 & 15.98 &    & 1.14 & 0.88 & (3) \\
12466 &  296.17462 & -14.78854 & 311 & 18.42 & 17.09 & 16.03 & 1.52 & 1.17 & 0.96 &  \\
40114 &  296.19916 & -14.86160 & 312 & 20.21 & 18.61 & 16.99 & 1.9  & 1.35 & 1.09 &  \\
11059 &  296.21259 & -14.80142 & 319 & 18.38 & 16.90 & 15.79 & 1.20 & 1.01 & 0.75 &  \\
20375 &  296.18277 & -14.74815 & 328 & 18.03 & 16.62 & 15.59 & 0.75 & 0.72 & 0.50 & C\\
11296 &  296.21729 & -14.77647 & 340 & 18.37 & 16.99 & 15.78 & 1.03 & 0.90 & 0.5  &  \\
30928 &  296.22229 & -14.90997 & 342 & 18.95 & 17.42 & 16.28 & 0.57 & 0.62 & 0.48 &  \\
20657 &  296.22934 & -14.65254 & 343 & 17.98 & 16.59 & 15.56 & 0.64 & 0.48 & 0.43 & C8.2(1)\\
13106 &  296.24048 & -14.84939 & 354 & 19.10 & 17.38 & 15.96 & 1.51 & 1.34 & 0.98 &  \\
20588 &  296.29919 & -14.69143 & 376 & 17.56 & 16.22 & 15.16 & 0.97 & 0.80 & 0.57 & C\\
11305 &  296.17957 & -14.77527 & 378 & 18.02 & 16.56 & 15.45 & 0.82 & 0.63 & 0.50 &  \\
30920 &  296.24246 & -14.91158 & 384 & 18.95 & 17.21 & 15.86 & 1.75 & 1.46 & 0.98 &  \\
40363 &  296.23649 & -14.85697 & 398 & 19.18 & 17.73 & 16.32 & 1.00 & 0.91 & 0.85 & C\\
11140 &  296.22876 & -14.79337 & 405 & 18.75 & 17.25 & 15.93 & 0.91 & 0.95 & 0.83 &  \\
20439 &  296.24640 & -14.73473 & 430 & 19.15 & 17.32 & 15.77 & 1.05 & 0.94 & 0.85 &  \\
10753 &  296.24295 & -14.83836 & 432 & 18.40 & 17.06 & 15.88 & 0.66 & 0.75 & 0.76 &  \\
40520 &  296.26687 & -14.74039 & 432 & 18.41 & 16.86 & 15.51 & 1.25 & 0.91 & 0.74 &  \\
31168 &  296.23883 & -14.87026 & 434 & 19.49 & 17.50 & 15.82 & 2.05 & 1.40 & 1.01 &  \\
21671 &  296.26569 & -14.72024 & 436 & 19.78 & 17.78 & 16.16 & 1.59 & 1.22 & 0.95 &  \\
11174 &  296.21774 & -14.78943 & 440 & 19.12 & 17.42 & 15.90 & 1.31 & 1.24 & 0.89 &  \\
20569 &  296.29114 & -14.69770 & 454 & 19.24 & 17.62 & 16.03 & 1.45 & 1.40 & 2.15 & \\
12445 &  296.21661 & -14.79057 & 454 & 20.26 & 18.34 & 16.43 & 2.42 & 1.78 & 1.25 & \\
21141 &  296.29965 & -14.69757 & 456 & 18.16 & 16.62 & 15.42 & 1.82 & 1.66 & 1.23 & \\
21234 &  296.28125 & -14.67393 & 466 & 20.16 & 18.21 & 16.25 & 1.54 & 1.41 & 1.25 & \\
12147 &  296.28897 & -14.81087 & 475 & 19.93 & 17.89 & 16.24 & 1.61 & 1.59 & 1.26 & \\
11299 &  296.25732 & -14.77645 & 494 & 19.06 & 17.30 & 15.73 & 2.16 & 2.07 & 1.67 & \\
13293 &  296.25696 & -14.83051 & 495 & 20.30 & 18.37 & 16.42 & 1.25 & 1.13 & 0.95 & \\
21029 &  296.19858 & -14.71837 & 501 & 19.11 & 17.07 & 15.55 & 1.5  & 1.1  & 0.8 &(5) \\
40102 &  296.19682 & -14.75119 & 526 & 19.69 & 17.43 & 15.69 & 2.12 & 1.31 & 1.05 & \\
12177 &  296.28424 & -14.80892 & 590 & 20.50 & 18.10 & 16.10 & 1.52 & 1.25 & 1.09 & \\
10807 &  296.21631 & -14.83196 & 747 & 20.37 & 18.00 & 15.89 & 1.92 & 2.13 & 1.59 & \\
40623 &  296.29793 & -14.74687 & 897 & 20.40 & 18.08 & 16.11 & 1.76 & 1.77 & 1.45 & \\
30268 &  296.25891 & -14.88796 & 998 & 16.73 & 15.44 & 14.45 & 1.35 & 1.47 & 1.21 & \\
\hline

\end{tabular}
\end{center}
* Low $K_S$ amplitude ($<0.4$ mag) for a Mira.\\
(1) Spectral type from Kacharov et al. (2012).\\
(2) N20558 was not used in the PL analysis as its image appears 
blended.\\
(3) N20840 has only 4 good observations at $J$, therefore its mean is uncertain and its amplitude
unknown.\\
(4) N30583 has seven observations only; P taken from Battinelli and Demers
(2011).  \\
(5) N21029 has a long-term trend (see Fig.~A2), mean given here for bright
cycle.
\label{tab_c} 
\end{table*}

\begin{table*}
\caption[]{Large Amplitude Variables.}
\begin{center}
\begin{tabular}{ccccccccccrrrc}
\hline
RA & Dec & N &  $J$ & $\delta J$& $H$ & $\delta  H$& $K_S$ & $\delta  K$ &  $J-K_S$
& NJ & NH & NK & note\\
\multicolumn{2}{c}{(2000.0)} && \multicolumn{7}{c}{(mag)} \\
\hline
 296.29922& -14.83650& 10293& 17.87& 0.88& 16.35& 0.72& 14.98& 0.07& 2.88& 16& 17&  8&\\
 296.19168& -14.82965& 10310& 18.32& 0.31& 16.80& 0.19& 15.67& 0.10& 2.65& 17& 15& 13&\\
 296.28415& -14.81194& 10371& 17.33& 0.16& 15.94& 0.12& 15.04& 0.07& 2.29& 15& 14& 12&\\
 296.23456& -14.77330& 10501& 18.00& 0.37& 16.56& 0.32& 15.48& 0.27& 2.52& 18& 18& 18&\\
 296.20154& -14.81282& 10968& 19.18& 0.60& 17.30& 0.30& 15.91& 0.20& 3.27& 18& 18& 18&\\
 296.18323& -14.76308& 11391& 18.39& 0.32& 16.94& 0.18& 15.82& 0.14& 2.57& 15& 16& 16&\\
 296.27222& -14.76143& 11401& 18.09& 0.24& 16.73& 0.17& 15.62& 0.07& 2.46& 17& 18& 13&\\
 296.22778& -14.76111& 11403& 18.35& 0.34& 17.01& 0.15& 16.18& 0.09& 2.17& 17& 14& 14&\\
 296.27774& -14.82287& 11991& 18.46& 0.18& 17.17& 0.22& 16.13& 0.20& 2.33& 18& 18& 18&\\
 296.24905& -14.81619& 12070& 17.49& 0.09& 16.92& 0.21& 15.98& 0.27& 1.51& 17& 18& 18&1\\
 296.25348& -14.76936& 12660& 18.38& 0.35& 16.83& 0.30& 15.54& 0.27& 2.84& 18& 18& 18&\\
 296.24728& -14.76419& 12711& 18.94& 1.05& 17.79& 0.92& 16.80& 0.63& 2.14& 18& 18& 17&\\
 296.24246& -14.82114& 13390& 19.99& 0.74& 18.40& 0.44& 16.77& 0.21& 3.23& 15& 15& 14&\\
 296.17993& -14.75767& 14105& 19.23& 0.39& 17.44& 0.33& 15.98& 0.26& 3.25& 17& 16& 16&\\
 296.27051& -14.73481& 20438& 19.39& 0.38& 17.70& 0.27& 16.08& 0.21& 3.32& 14& 15& 15&\\
 296.18378& -14.68249& 20608& 19.57& 1.40& 17.64& 0.95& 16.06& 0.67& 3.51& 15& 17& 17&\\
 296.26770& -14.68061& 20614& 19.60& 1.62& 17.60& 0.76& 16.30& 0.53& 3.30& 16& 17& 17&\\
 296.17764& -14.64240& 21316& 18.63& 0.79& 17.47& 0.64& 16.42& 0.45& 2.20& 15& 17& 17&\\
 296.26511& -14.96155& 30767& 18.07& 0.40& 17.09& 0.30& 16.45& 0.20& 1.62& 14& 15& 14&\\
 296.21112& -14.91074& 30924& 19.46& 0.36& 17.55& 0.20& 16.25& 0.13& 3.21& 13& 14& 12&\\
 296.30179& -14.90356& 30961& 19.75& 0.26& 17.75& 0.18& 15.96& 0.44& 3.78&  6&  6&  6&\\
 296.18011& -14.75091& 40030& 17.02& 0.16& 16.27& 0.20& 15.81& 0.12& 1.21& 23& 30& 29&\\
 296.22681& -14.75065& 40275& 17.13& 0.34& 15.92& 0.29& 15.11& 0.28& 2.01& 34& 34& 34&\\
 296.23346& -14.74996& 40327& 17.71& 0.18& 16.93& 0.23& 16.52& 0.18& 1.19& 31& 35& 33&\\
 296.24622& -14.86655& 40419& 17.98& 0.08& 16.67& 0.06& 15.79& 0.08& 2.19& 20& 18& 20&2\\
 296.26105& -14.73764& 40493& 18.29& 0.15& 16.78& 0.16& 15.38& 0.19& 2.91& 22& 21& 28&\\
 296.27310& -14.75146& 40538& 17.31& 0.16& 16.57& 0.17& 15.96& 0.20& 1.35& 29& 30& 31&\\
\hline
\end{tabular}
\end{center}
\label{tab_var1} 
(1) N12070 is probably a Mira, with $\Delta K_S < 0.6$ mag, and possible
periods of around 545 or 215 days, but its image is confused at shorter
wavelengths.\\ (2) N40419 is a Mira with a period of 193 days, but 
its photometry is contaminated by nearby sources.
\end{table*}

\begin{center}
\onecolumn
\begin{longtable}{cccccccccccccc}
\caption[Small Amplitude Variables.]{Small Amplitude Variables.} \label{tab_var2} \\
\hline
RA & Dec & N & $J$& $\delta J$& $H$ & $\delta  H$& $K_S$ & $\delta  K$ &$J-K_S$
& NJ & NH & NK&\\
\multicolumn{2}{c}{(2000.0)} && \multicolumn{7}{c}{(mag)} \\
\hline
\endfirsthead

\hline
RA & Dec & N & $J$& $\delta J$& $H$ & $\delta  H$& $K_S$ & $\delta  K$ &$J-K_S$
& NJ & NH & NK&\\
\multicolumn{2}{c}{(2000.0)} && \multicolumn{7}{c}{(mag)} \\
\hline
\endhead
 \multicolumn{12}{l}{{Continued on Next Page\ldots}} \\
\endfoot
\endlastfoot
 296.22696& -14.80191& 10032& 14.48& 0.05& 13.66& 0.03& 13.34& 0.05& 1.14& 14& 14& 16\\
 296.24005& -14.80796& 10074& 14.74& 0.09& 13.94& 0.10& 13.66& 0.10& 1.08& 18& 18& 18\\
 296.21835& -14.80118& 10077& 15.94& 0.11& 14.97& 0.12& 14.55& 0.11& 1.40& 18& 18& 18\\
 296.26028& -14.81812& 10152& 16.32& 0.17& 15.26& 0.14& 14.71& 0.11& 1.61& 18& 18& 18\\
 296.22321& -14.76671& 10200& 16.47& 0.08& 15.32& 0.06& 14.61& 0.06& 1.86& 17& 16& 17\\
 296.19437& -14.82407& 10330& 17.33& 0.14& 16.16& 0.13& 15.52& 0.11& 1.82& 18& 18& 18\\
 296.17441& -14.81870& 10343& 17.35& 0.11& 16.12& 0.13& 15.44& 0.11& 1.92& 15& 18& 18\\
 296.23013& -14.81558& 10356& 17.00& 0.16& 15.84& 0.16& 15.20& 0.14& 1.80& 18& 18& 18\\
 296.23392& -14.80649& 10400& 16.97& 0.11& 15.74& 0.09& 15.00& 0.07& 1.97& 15& 16& 18\\
 296.24081& -14.80368& 10408& 16.93& 0.13& 15.77& 0.09& 15.11& 0.07& 1.82& 17& 17& 17\\
 296.19797& -14.80259& 10411& 16.92& 0.18& 15.82& 0.14& 15.23& 0.10& 1.69& 18& 18& 18\\
 296.29706& -14.80268& 10412& 17.34& 0.17& 16.08& 0.14& 15.18& 0.13& 2.17& 17& 16& 18\\
 296.22131& -14.79940& 10425& 17.30& 0.18& 16.11& 0.13& 15.40& 0.10& 1.90& 18& 18& 18\\
 296.28729& -14.79597& 10433& 17.46& 0.18& 16.30& 0.14& 15.55& 0.08& 1.91& 18& 18& 16\\
 296.20374& -14.79388& 10439& 17.33& 0.16& 16.08& 0.11& 15.37& 0.08& 1.96& 18& 17& 17\\
 296.28287& -14.78797& 10460& 17.32& 0.26& 16.24& 0.20& 15.66& 0.15& 1.66& 18& 18& 18\\
 296.22369& -14.83960& 10743& 17.57& 0.29& 16.43& 0.24& 15.67& 0.19& 1.90& 18& 18& 18\\
 296.27734& -14.83831& 10755& 17.42& 0.20& 16.23& 0.16& 15.52& 0.13& 1.91& 18& 18& 18\\
 296.27213& -14.83158& 10809& 18.07& 0.14& 16.90& 0.19& 16.13& 0.19& 1.93& 17& 18& 18\\
 296.24747& -14.82692& 10839& 18.15& 0.09& 16.63& 0.09& 15.41& 0.07& 2.74& 14& 15& 15\\
 296.29504& -14.82570& 10850& 17.89& 0.38& 16.68& 0.30& 15.90& 0.18& 1.99& 18& 18& 18\\
 296.19476& -14.82463& 10859& 17.67& 0.21& 16.57& 0.17& 15.91& 0.14& 1.75& 18& 17& 18\\
 296.24554& -14.82244& 10876& 17.71& 0.26& 16.41& 0.20& 15.50& 0.12& 2.21& 18& 18& 18\\
 296.20242& -14.81780& 10917& 17.35& 0.08& 16.07& 0.09& 15.24& 0.07& 2.11& 14& 16& 16\\
 296.28149& -14.81671& 10935& 17.58& 0.10& 16.60& 0.11& 16.16& 0.17& 1.42& 14& 15& 18\\
 296.27066& -14.80596& 11021& 17.54& 0.24& 16.41& 0.19& 15.72& 0.12& 1.82& 18& 18& 17\\
 296.24573& -14.79360& 11139& 17.98& 0.31& 16.72& 0.11& 15.67& 0.12& 2.31& 17& 14& 17\\
 296.24271& -14.78809& 11187& 17.90& 0.08& 16.50& 0.06& 15.68& 0.07& 2.22& 16& 17& 18\\
 296.27832& -14.78018& 11271& 17.42& 0.24& 16.25& 0.21& 15.55& 0.15& 1.87& 18& 18& 18\\
 296.17908& -14.77972& 11273& 17.63& 0.21& 16.43& 0.14& 15.71& 0.07& 1.92& 18& 16& 14\\
 296.18372& -14.77252& 11335& 17.51& 0.19& 16.28& 0.18& 15.41& 0.12& 2.10& 18& 18& 18\\
 296.20221& -14.76910& 11364& 17.59& 0.06& 16.26& 0.03& 15.38& 0.03& 2.21& 15& 13& 17\\
 296.22644& -14.76749& 11372& 17.53& 0.31& 16.46& 0.22& 15.70& 0.13& 1.83& 18& 18& 18\\
 296.19781& -14.76344& 11389& 17.58& 0.23& 16.45& 0.18& 15.64& 0.14& 1.93& 18& 18& 17\\
 296.28235& -14.76346& 11392& 18.41& 0.19& 17.01& 0.08& 15.96& 0.07& 2.46& 17& 16& 18\\
 296.24139& -14.84622& 11764& 18.82& 0.23& 17.52& 0.12& 16.51& 0.10& 2.31& 15& 15& 16\\
 296.29608& -14.84267& 11794& 17.73& 0.24& 16.83& 0.16& 16.36& 0.11& 1.37& 17& 18& 16\\
 296.24057& -14.79573& 12373& 18.57& 0.32& 17.52& 0.24& 16.80& 0.09& 1.77& 16& 17& 13\\
 296.18201& -14.78503& 12496& 18.17& 0.28& 16.81& 0.25& 15.74& 0.22& 2.44& 17& 18& 18\\
 296.25473& -14.75607& 12784& 18.51& 0.22& 17.03& 0.17& 15.89& 0.11& 2.62& 18& 18& 17\\
 296.20816& -14.72616& 20022& 14.09& 0.09& 13.21& 0.08& 12.83& 0.08& 1.26& 16& 16& 17\\
 296.26938& -14.66636& 20311& 17.19& 0.12& 16.20& 0.16& 15.49& 0.07& 1.69& 14& 17& 14\\
 296.23389& -14.72910& 20463& 17.86& 0.16& 16.87& 0.09& 16.51& 0.14& 1.36& 11& 13& 13\\
 296.29358& -14.72387& 20496& 17.36& 0.20& 16.28& 0.17& 15.65& 0.13& 1.71& 16& 17& 17\\
 296.17297& -14.71139& 20539& 17.96& 0.22& 16.59& 0.19& 15.50& 0.10& 2.46& 13& 17& 15\\
 296.26337& -14.70901& 20547& 17.34& 0.09& 16.23& 0.09& 15.66& 0.08& 1.68& 15& 16& 17\\
 296.24963& -14.71943& 21021& 18.72& 0.09& 17.38& 0.05& 16.42& 0.06& 2.30& 14& 15& 16\\
 296.28815& -14.67794& 21217& 17.67& 0.16& 16.75& 0.13& 16.20& 0.13& 1.47& 15& 16& 16\\
 296.25790& -14.90380& 30244& 17.38& 0.14& 16.17& 0.12& 15.44& 0.10& 1.95& 14& 15& 14\\
 296.20166& -14.88640& 30271& 16.68& 0.11& 15.57& 0.10& 14.93& 0.08& 1.75& 13& 14& 14\\
 296.28564& -14.87809& 30285& 16.97& 0.16& 15.88& 0.13& 15.18& 0.04& 1.79& 14& 15& 10\\
 296.20511& -14.87565& 30590& 17.68& 0.28& 16.48& 0.05& 16.01& 0.12& 1.67& 14& 10& 14\\
 296.19254& -14.86788& 30611& 18.29& 0.26& 16.99& 0.20& 16.08& 0.13& 2.21& 14& 15& 14\\
 296.29318& -14.90858& 30934& 18.03& 0.09& 16.89& 0.05& 15.53& 0.04& 2.49& 14& 13& 11\\
 296.17282& -14.75015& 40002& 17.59& 0.11& 16.53& 0.06& 15.91& 0.04& 1.68& 26& 26& 24\\
 296.17505& -14.85986& 40007& 17.97& 0.11& 16.70& 0.12& 15.75& 0.04& 2.22& 24& 30& 22\\
 296.17966& -14.73710& 40026& 17.49& 0.19& 16.36& 0.15& 15.59& 0.10& 1.90& 16& 31& 27\\
 296.20724& -14.74512& 40155& 17.08& 0.09& 16.00& 0.06& 15.40& 0.05& 1.68& 34& 35& 29\\
 296.22379& -14.73659& 40257& 16.01& 0.02& 15.20& 0.05& 14.95& 0.03& 1.06& 18& 21& 17\\
 296.22452& -14.86319& 40261& 17.48& 0.17& 16.42& 0.12& 15.89& 0.10& 1.60& 30& 30& 29\\
 296.24301& -14.74652& 40397& 14.10& 0.07& 13.30& 0.06& 12.99& 0.07& 1.11& 34& 36& 36\\
 296.24622& -14.86655& 40419& 17.98& 0.08& 16.67& 0.06& 15.79& 0.08& 2.19& 20& 18& 20\\
 296.24634& -14.86488& 40421& 17.41& 0.06& 16.40& 0.05& 15.82& 0.04& 1.59& 25& 24& 21\\
 296.25229& -14.74336& 40445& 17.70& 0.18& 16.57& 0.18& 15.80& 0.08& 1.90& 31& 35& 29\\
 296.25253& -14.86060& 40446& 18.61& 0.41& 17.57& 0.43& 16.75& 0.31& 1.86& 33& 33& 29\\
 296.25864& -14.73680& 40476& 17.71& 0.08& 16.44& 0.05& 15.53& 0.03& 2.18& 29& 25& 23\\
 296.25992& -14.74793& 40486& 18.59& 0.17& 17.73& 0.29& 16.70& 0.13& 1.88& 26& 33& 25\\
 296.26254& -14.84794& 40501& 17.26& 0.14& 16.14& 0.09& 15.51& 0.11& 1.76& 29& 25& 28\\
 296.27545& -14.85393& 40547& 17.45& 0.17& 16.30& 0.10& 15.66& 0.10& 1.79& 26& 25& 32\\
\hline
\end{longtable}
\end{center}
\twocolumn

\subsection{Comparison with Battinelli and Demers (2011)}

Battinelli \& Demers (2011) discuss long period variables discovered in
their 32 arcmin square survey of NGC\,6822 using 1.5-m and 1.6-m telescopes,
over a period of somewhat over three years.  There is considerable overlap
with our work although their survey extends further to the east (they also
have unpublished data extending to the west).  Twenty of their variables
fall within the area we surveyed.

Table \ref{tab_bd} lists 16 of those 20 variables that we also regard as
large amplitude variables and for which we derived periods.  It includes the
long period Cepheid, our N10170.  The other 4 are briefly described below:\\
 BD~v13, for which they find a period of 466 days, corresponds to our N20438
(Table~\ref{tab_var1}) and its $K$ light curve is illustrated in
Fig.~\ref{fig_long1}.  Although it shows large amplitude variations its
behaviour was not sufficiently regular for us to identify it as periodic,
and it is therefore included in Table~\ref{tab_var1}.  However, knowing the
period and examining the light curve it is possible to see that it underwent
two maxima during the time we observed it, the first at $K_S\sim 15.8$ mag
and the second at $K_S\sim 15.3$ mag.\\
 BD~v14 is N40538 and is also a variable (Table~\ref{tab_var2}), but without
a clearly defined period.\\  
BD~v9 is N20287 which is not obviously significantly variable.\\ 
 BD~v21 was very faint on the $H$ frame that we used as a reference, and was
therefore not extracted in our survey, although it is clearly seen on
the $K$ frames.  Battinelli \& Demers determined P=613 days and
$\Delta K_S=0.7$ mag for this star.

\begin{figure}
\includegraphics[width=8.5cm]{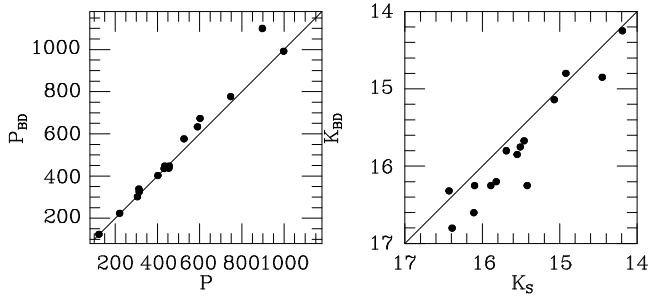}
\caption{A comparison of the periods and mean $K$ magnitudes derived here and by
Battinelli \& Demers (2011).}
\label{fig_bp}
\end{figure}

\begin{figure}
\begin{center}
\includegraphics[width=6cm]{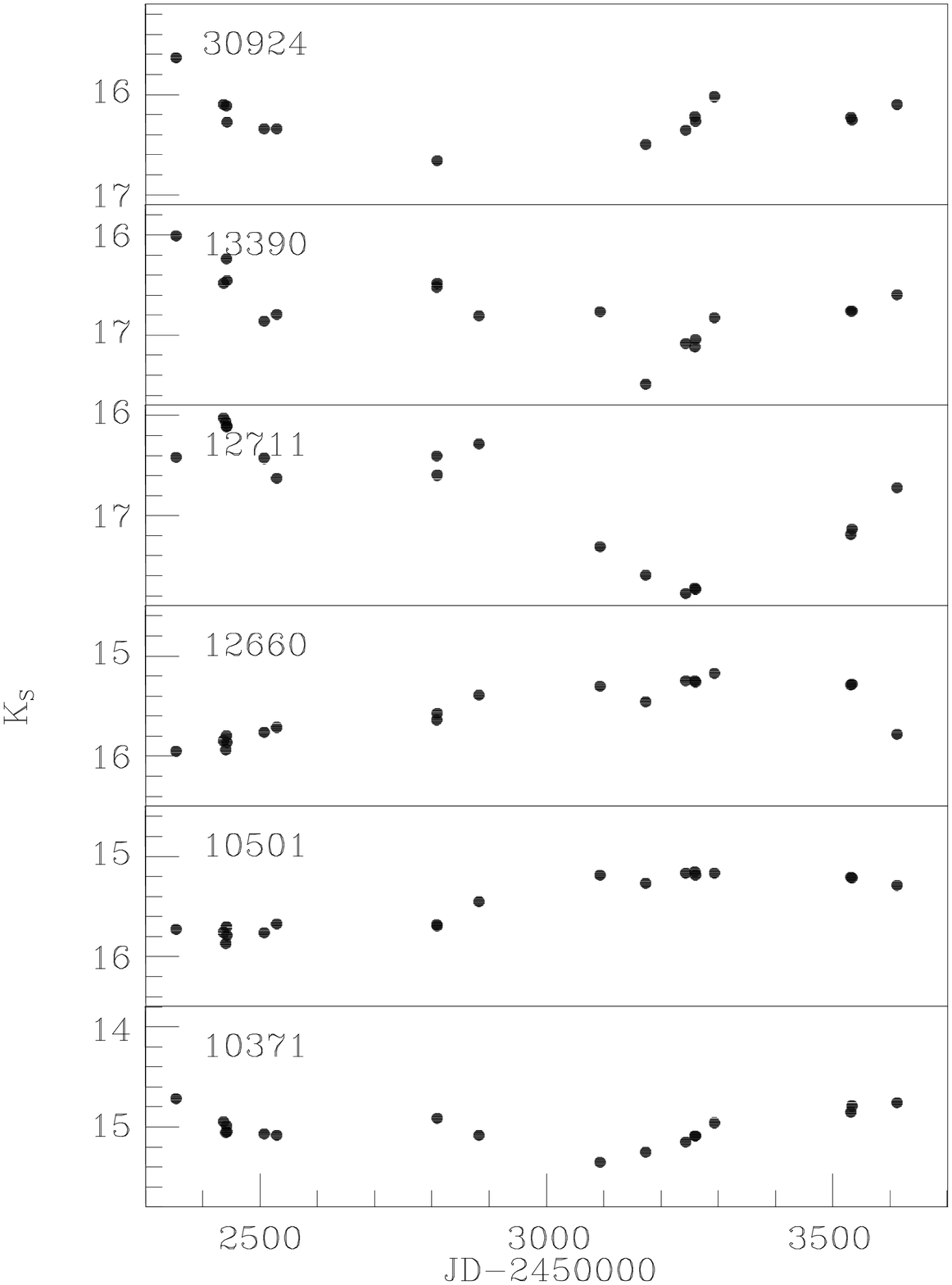}
\includegraphics[width=6cm]{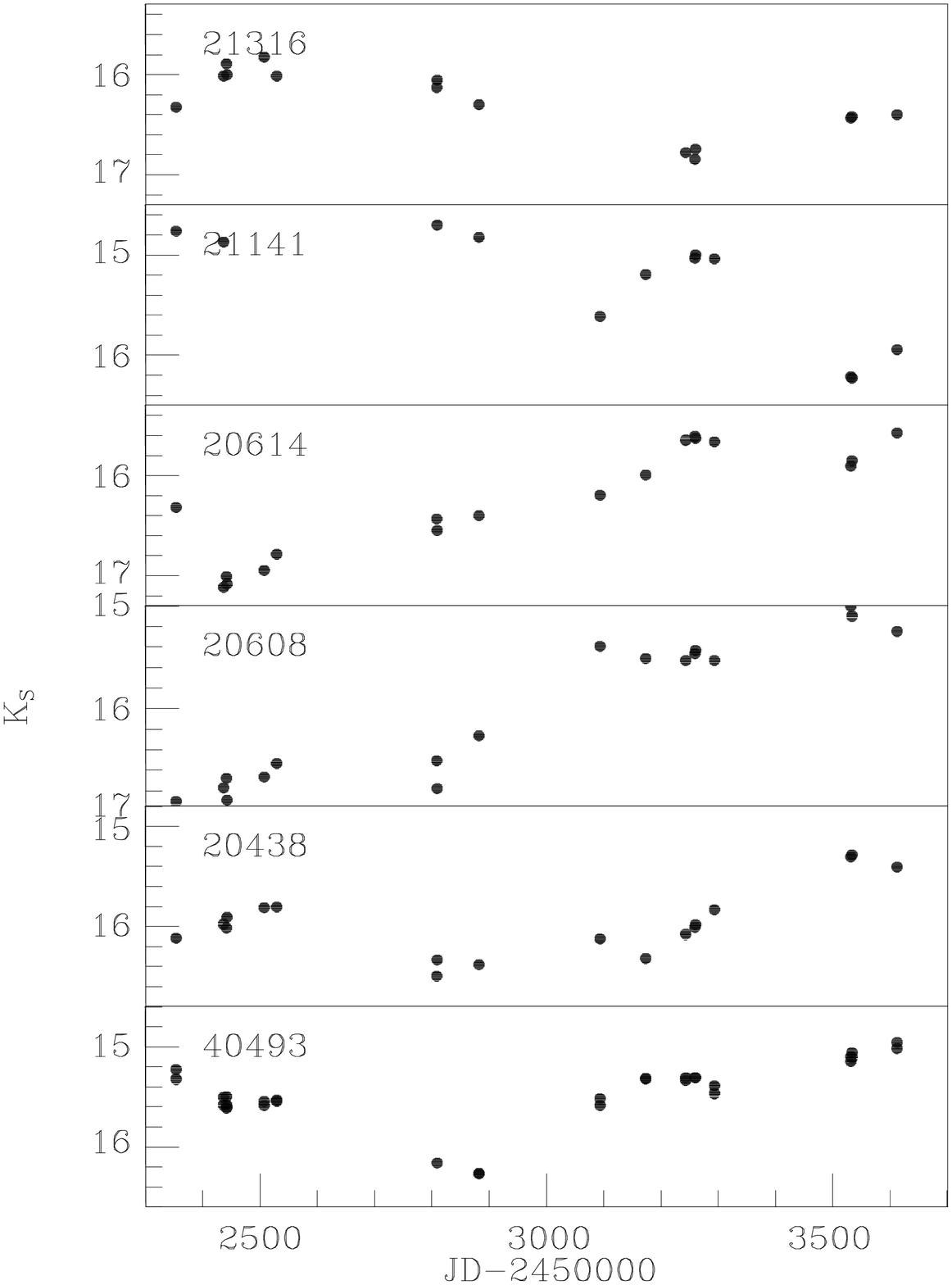}
\caption{The light-curves for some of the large amplitude variables without
obvious periodicity.}
\label{fig_long1}
\end{center}
\end{figure}

All of the other variables from Battinelli \& Demers's table~3 are outside
the positional range of our survey.  Fig.~\ref{fig_bp} compares their
periods and mean $K_S$ magnitudes with ours for the stars in common.  We
note that while the periods are in reasonable agreement there does appear to
be a systematic difference in the mean $K_S$ magnitude.  The mean difference is
0.25 mag, or 0.21 mag leaving out BD~v19=N21141 where the mean magnitudes
differed by 0.8 mag.  It is disturbing to find such a large difference in
the magnitudes of potential distance indicators and the matter is worth
further investigation.  There are no non-variable stars in Battinelli \&
Demers to compare with ours, but we have made the comparison with the UKIRT
photometry of Sibbons et al.  (2012).  The difference between their $K_S$
magnitudes and ours, both uncorrected for interstellar reddening, is only
0.05 mag at $K_S=16$ mag and about 0.1 mag at $K_S=17$ mag.  At the fainter
magnitudes crowding can be a problem for matching objects between the two
catalogues, quite apart from possible photometric difficulties.  Neither
study has taken account of any possible colour equation, both being
essentially on the natural system.  Though colour effects are probably not
significant if $J-K_S<1.0$ mag, they may need to be considered for the reddest
stars considered here, but it is extremely difficult to do the calibration
work required to quantify the effect.

Given that our entire field falls within the area discussed by Battinelli \&
Demers it is a little surprising that they do not find more of the Mira
variables that we identify.  Six more of them are to be found in their
table~4 which lists SR and irregular variables and Table~\ref{tab_bd2} cross
references our identification numbers with theirs (note that their BD~v116
has the identical coordinates to their BD~v12).  They do not appear to have
found the other 39.

Among their SR variables  BD~v109 is not measurable on our image where it is
extended, while BD~v117 is another very red object, visible at $K_S$ but not
at shorter wavelengths.  The other variables they list are outside of our
survey area.\\

\begin{table}
\caption[]{Variables in common with Battinelli \& Demers (2011) table 3}
\begin{center}
\begin{tabular}{cccccc}
\hline
N & P & $K_S$ & BD & P & $K_S$\\
& (days) & (mag) & & (days) & (mag) \\
\hline
40102 & 526 & 15.69 & 1 & 576 & 15.80\\
40114 & 312 & 16.99 & 2 & 339 & 17.25\\
20331 & 314 & 15.46 & 3 & 326 & 15.67\\
20134 & 402 & 15.07 & 4 & 403 & 15.14\\
10807 & 747 & 15.89 & 5 & 777 & 16.25\\
12445 & 454 & 16.43 & 6 & 437 & 16.32\\
31168 & 434 & 15.82 & 7 & 447 & 16.20\\
10198 & 602 & 14.19 & 8 & 673 & 14.25\\
10170 & 123 & 14.92 & 10& 124 & 14.80\\
30268 & 998 & 14.45 & 11& 992 & 14.85\\
40520 & 432 & 15.51 & 12& 436 & 15.75\\
12177 & 590 & 16.10 & 15& 633 & 16.25\\
40590 & 221 & 16.39 & 16& 223 & 16.80\\
40623 & 897 & 16.11 & 17& 1100& 16.60\\ 
21141 & 456 & 15.42 & 19& 448 & 16.25\\
30583 & 305: & 15.55 & 20& 302 & 15.85\\
\hline
\end{tabular}
\end{center}
\label{tab_bd} 
\end{table}

\begin{table}
\caption[]{Variables in common with Battinelli \& Demers (2011) table 4}
\begin{center}
\begin{tabular}{ccl}
\hline
BD &  N  & remark\\
\hline
100 & 30981\\
101 & 11173\\
102 & 11296 & Mira Table~\ref{tab_c}\\
103 & 11414\\
104 & 20468\\
105 &20239 & Mira Table~\ref{tab_c}\\
106 & 11372 \\
107 & 10501 & LPV trend Table~\ref{tab_var1}\\
108 & 30920 & Mira Table~\ref{tab_c}\\
111 & 20892 \\
112 & 12660 & LPV trend Table~\ref{tab_var1}\\ 
113 & 11299 & Mira Table~\ref{tab_c}\\
114 & 11362\\
115 & 40482\\
116 & 40520 & =ID12 Mira Table~\ref{tab_c}\\
118 & 20569 & Mira Table~\ref{tab_c}\\
119 & 20588 & Mira Table~\ref{tab_c}\\
\hline
\end{tabular}
\end{center}
\label{tab_bd2} 
\end{table}

\subsection{M-type Miras}

The most luminous of the stars in Table~\ref{tab_o}(a), N20004, is a known
variable, NGC6822\,V12.  Kayser (1967) describes this as a semi-regular and
writes `Most of the time it varies regularly with a period of 640 days, but
every three to six cycles it does something else for 180 days'.  Note that
640 days is significantly different from the 854 days that we found. Massey
(1998) gives its spectral type as M2.5-3I: and $JHK$ photometry was
published by Elias \& Frogel (1985).  Its amplitude is lower, and luminosity
higher, than those of the other variables considered here and it is possibly
the descendent of a more massive star and different from the other O-rich
variables, but that is not entirely clear.  Its luminosity is comparable to
the supergiants discussed by Levesque \& Massey (2012).

N10198 was identified as a variable, v198, by Baldacci et al. (2005).
N10184 and N11226 were identified as variables, v1838 and v1534,
respectively, by Antonello et al. (2002). N40139 appears in the GCVS as v16.

The bolometric magnitudes of the presumed O-rich stars
were calculated by fitting a blackbody to the $JHK$ fluxes, following the
procedure used by Robertson \& Feast (1981) and by Feast et al.  (1989).  In
practice, for these stars with very thin shells, bolometric magnitudes
derived in this way differ insignificantly ($<0.03$ mag) from those
calculated using the bolometric corrections defined for the C~stars (section
7).  The results are listed in Table~\ref{tab_obol} and illustrated in a PL
relation (Fig.~\ref{fig_PL}).

The two short period stars are presumably similar to the short period O-rich
Miras found in globular clusters (Feast et al.  2002; Whitelock et al. 
2008), and their luminosities are comparable to those of the short period
C-rich Miras.  The brighter, longer period, stars appear to represent a
somewhat younger population.  They are probably similar to long-period
O-rich Miras that are found in the LMC, many of which have s-process
enhancements (Lundgren 1988; Smith et al.  1995).  They are considered
further in the next section.

\begin{table}
v\caption[]{Bolometric magnitudes of the assumed O-rich Miras. }
\begin{center}
\begin{tabular}{cccc}
\hline
N & P & $m_{bol}$ & $(J-K_S)_0$\\
& (days) & \multicolumn{2}{c}{(mag)}\\
\hline
 12557 &  158 &  20.13 &  1.05\\ 
 11226 &  257 &  19.27 &  1.37\\ 
 20331 &  314 &  18.51 &  1.06\\ 
 10184 &  370 &  18.27 &  1.25\\ 
 30133 &  401 &  18.19 &  1.15\\ 
 20134 &  402 &  18.21 &  1.23\\ 
 40139 &  545 &  17.05 &  1.28\\ 
 10198 &  602 &  17.36 &  1.27\\ 
 30292 &  637 &  17.80 &  1.15\\ 
 10091 &  638 &  17.34 &  1.24\\ 
 20004 &  854 &  15.62 &  1.10\\ 
\hline
\end{tabular}
\end{center}
\label{tab_obol} 
\end{table}
\begin{table}
\caption[]{Bolometric magnitudes of the assumed C-rich Miras.}
\begin{center}
\begin{tabular}{cccc}
\hline
N & P & $m_{bol}$ & $(J-K_S)_0$\\
& (days) & \multicolumn{2}{c}{(mag)}\\
\hline
 12790 & 182 & 19.82 &   1.34\\
 10817 & 214 & 19.41 &   1.56\\
 20540 & 223 & 19.63 &   1.39\\
 40590 & 223 & 19.74 &   1.52\\
 12751 & 231 & 19.62 &   1.50\\
 11032 & 239 & 19.34 &   2.16\\
 10748 & 243 & 19.48 &   2.36\\
 20578 & 246 & 19.50 &   1.56\\
 20542 & 255 & 19.42 &   1.72\\
 11226 & 257 & 19.27 &   1.31\\
 30430 & 269 & 19.25 &   1.71\\
 12208 & 278 & 19.02 &   2.15\\
 21419 & 278 & 19.29 &   3.26\\
 13364 & 286 & 19.26 &   2.14\\
 12400 & 301 & 19.30 &   2.02\\
 30583 & 302 & 19.19 &   2.39\\
 20558 & 304 & 18.47 &   2.80\\
 20239 & 304 & 19.04 &   2.27\\
 20840 & 306 & 19.24 &   2.62\\
 12466 & 311 & 19.40 &   2.29\\
 20331 & 314 & 18.51 &   1.02\\
 40114 & 316 & 19.76 &   3.16\\
 11059 & 319 & 19.11 &   2.49\\
 20375 & 328 & 18.96 &   2.34\\
 11296 & 340 & 19.06 &   2.49\\
 30928 & 342 & 19.58 &   2.57\\
 20657 & 343 & 18.93 &   2.32\\
 13106 & 354 & 18.98 &   3.04\\
 10184 & 370 & 18.27 &   1.20\\
 20588 & 376 & 18.53 &   2.30\\
 11305 & 378 & 18.78 &   2.47\\
 30920 & 384 & 18.94 &   2.99\\
 40363 & 398 & 19.41 &   2.77\\
 30133 & 401 & 18.18 &   1.11\\
 20134 & 402 & 18.22 &   1.18\\
 11140 & 405 & 19.10 &   2.72\\
 20439 & 430 & 18.61 &   3.28\\
 40520 & 432 & 18.64 &   2.80\\
 10753 & 432 & 19.19 &   2.42\\
 31168 & 434 & 18.44 &   3.57\\
 21671 & 436 & 18.85 &   3.52\\
 11174 & 440 & 18.82 &   3.12\\
 12445 & 454 & 18.76 &   3.73\\
 20569 & 454 & 18.89 &   3.11\\
 21141 & 456 & 18.68 &   2.64\\
 21234 & 466 & 18.50 &   3.81\\
 12147 & 475 & 18.88 &   3.59\\
 11299 & 494 & 18.57 &   3.23\\
 13293 & 495 & 18.69 &   3.78\\
 21029 & 501 & 18.34 &   3.46\\
 40102 & 526 & 18.11 &   3.90\\
 40139 & 545 & 17.03 &   1.23\\
 12177 & 590 & 18.08 &   4.30\\
 10198 & 602 & 17.36 &   1.21\\
 30292 & 637 & 17.81 &   1.10\\
 10091 & 638 & 17.34 &   1.19\\
 10807 & 747 & 17.71 &   4.38\\
 20004 & 854 & 15.60 &   1.06\\
 40623 & 897 & 18.07 &   4.39\\
 30268 & 998 & 17.84 &   2.18\\
\hline
\end{tabular}
\end{center}
\label{tab_cbol} 
\end{table}

\subsection{C-type Miras}

These are discussed below in section~7.

\section{Completeness of large amplitude variable survey}
We measured pulsation periods for 61 Mira variables in our survey
area, compared to the 20 measured by Battinelli \& Demers (2011) in the same
area (only 16 of these are in common).  So there is no question that we
significantly improved on their count.  However, we note that our survey
will still be incomplete for the following reasons:\\
 (1) 
We may have missed a very small number of very red, dust enshrouded,
large amplitude variables
entirely (see section 5.1) and indeed we did miss one of those found by
Battinelli \& Demers.
Nevertheless it is interesting to
see that we have one assumed C-rich Mira, with a period of 998 days (N30268
for which there is no measured spectral type). 

%IS THERE AN EARLIER PAPER BY PAW ON THIS? - possibly) 
 It has been suggested that at periods longer than about 1000 days the stars
are sufficiently massive for hot bottom burning to occur and that will
result in O-, rather than C-rich Miras (Feast 2009).  The longest period
C-rich Miras in the LMC are also just under 1000 days (Whitelock et al. 
2003).  It would obviously be very interesting to know if we have any OH/IR
stars in NGC\,6822 with periods over 1000 days.\\
 (2) Some Miras behave erratically and very long-term monitoring is 
necessary to characterize them. That is one of the reasons for differences
with Battinelli \& Demers. Most such stars will appear in
Tables~\ref{tab_var1} or \ref{tab_var2}.\\
 (3) Confusion, especially in the crowded inner regions, limits our
ability to measure the Miras, particularly at $J$.

All of these factors are relevant, but the only one that is likely to
seriously affect the total count is item (2).

\section{Period-Luminosity Relations}

There are a variety of ways in which bolometric magnitudes can be measured
or estimated, depending on the information available and, to some extent, on
the desired objective.  Given that one is almost always limited by temporal
and/or spectral coverage any chosen approach is a compromise.  Kerschbaum,
Lebzelter \& Makul (2010) discuss different approaches and show that they
can lead to very different results: over 0.5 mag spread at a particular
colour.  We also note that Kamath et al.  (2010) and Groenewegen et al. 
(2007) derive bolometric magnitudes for several variable stars in the SMC
cluster NGC\,419 using the same Spitzer data and slightly different $JHKL$
values.  Their bolometric magnitudes differ by amounts that range from --0.1
to 0.4 mag for the same star.
 
These uncertainties present difficulties when attempting to compare
bolometric luminosities with theoretical predictions.  For instance, in a
plot of bolometric magnitude against period (their fig.~7), Kamath et al. 
place a group of NGC\,419 carbon-rich semi-regular variables on their computed
fundamental sequence and a group of brighter shorter period variables on
their first overtone sequence.  This is contrary to the usually accepted
model of the overall evolution that increasing luminosity and period implies
decreasing mode.  Such a normal evolutionary sequence is supported by the 
$K-\log P$ plot for these same NGC419 variables. This places
them all together in a single group  on the first overtone sequence
in a ``Wood" PL diagram (for instance the LMC/SMC plots of Ita et al.\ 2004).

For the purpose of this paper we follow the same procedure for determining
bolometric magnitudes as in our previous papers as this will give us
consistent values that are good, at least, for estimating distances via the
PL relation.
  
 Bolometric magnitudes for the presumed C stars were calculated in the same
way as in our earlier papers (e.g.  Whitelock et al.  2009), by applying a
colour-dependent bolometric correction to the reddening-corrected $K$
magnitudes on the SAAO system.  The magnitudes given in Table~\ref{tab_o},
which are on the 2MASS system, are converted to the SAAO system following
Carpenter (2001 and web page
update\footnote{http://www.astro.caltech.edu/~jmc/2mass/v3/transformations/}).  The resulting bolometric magnitudes
are listed in Table~\ref{tab_cbol}.

Note that the bolometric magnitudes derived for stars with faint $J$
magnitudes are rather uncertain due to photometric errors and the increased
possibility of confusion.

\begin{figure}
\includegraphics[width=8.5cm]{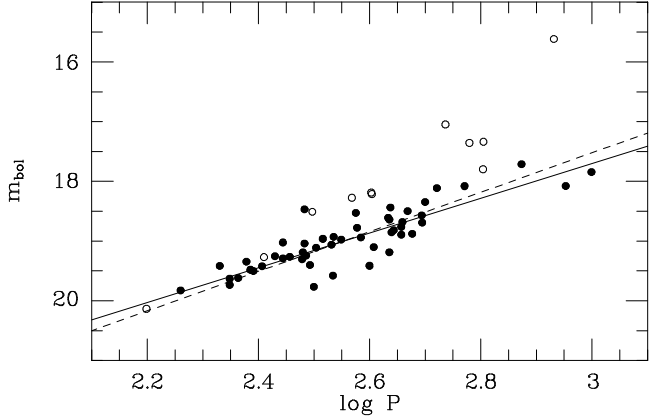}
\caption{Bolometric PL for the large amplitude variables in NGC\,6822; open
symbols are those from Table~\ref{tab_o}(a); closed
symbols represent those from Table~\ref{tab_c}(b). The solid line
is the best fit to the closed circles while the dashed line is the best
fitting one with the same slope as the LMC PL relation.}
\label{fig_PL}
\end{figure}

Figure \ref{fig_PL} shows a PL relation. 
A least squares fit to the 50 presumed C-rich Miras gives the following
result: \begin{equation}
m_{bol}=19.16(\pm0.04)-2.91(\pm0.22)[\log P -2.5], \end{equation}
with a scatter of 0.23 mag. This expression is shown as a solid line in
Fig.~\ref{fig_PL}.

Alternatively, if we assume that the slope of the PL relation is the same as
that found in the LMC, $-3.31\pm0.24$ (Whitelock et al.  2009), then we
derive a zero point of $19.18\pm0.03$, with a scatter of 0.24 mag for the
same 50 stars.  This line is also shown in Fig.~\ref{fig_PL}.

If we restrict the stars in NGC\,6822 to cover the same range of periods as
used to derive the LMC PL relation, i.e. $220 < P < 500 $, then the
zero-point is $19.20\pm 0.04$, with a scatter of 0.23 mag for 41 stars.

Leaving out the stars with the faintest $J$ magnitudes, $J>20.0$ mag, which
have the most uncertain bolometric magnitudes, we find a zero point of
$19.16\pm 0.04$, with a scatter of 0.22 mag for 40 stars.

Leaving out only star N40114, which is unusually red ($J-K\sim 3.28$) for a
P=313 day Mira and rather faint with respect to the PL relation, gives a zero
point of $19.17\pm 0.03$, with a scatter of 0.22 mag for 49 stars

Thus these results are consistent with the PL relation in NGC\,6822 having
the same slope as it does for the LMC.  If we assume that the distance of
the LMC is $(m-M)_0=18.50$ mag, the PL relation derived from LMC Miras is
$$ M_{bol}=-4.38-3.31[\log P -2.5],$$ and using the zero point of $19.18\pm
0.03$, the distance modulus for NGC\,6822 is $(m-M)_0=23.56 (\pm0.03)$ mag. 
The uncertainty does not include the uncertainty on the LMC distance.  This
can be compared with values of 23.40 mag and 23.49 derived from Cepheids and
RR Lyraes, respectively (Feast et al.  2012; Clementini et al.  2003).

\begin{figure} \includegraphics[width=8.5cm]{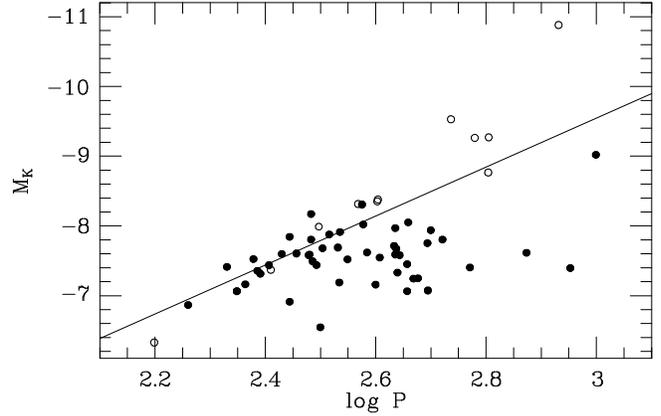} \caption{PL($K$)
for the large amplitude variables in NGC\,6822; symbols as in
Fig.~\ref{fig_PL}.  The solid line is the PL($K$) derived for the Galactic
O-rich Miras assuming $(m-M)_{LMC}=18.5$ mag. The absolute $K$ magnitudes 
shown here assumes $(m-M)_{N6822}=
23.38$ mag, as was determined from the 4 faintest O-rich Miras.}
\label{fig_PLK} \end{figure}

In the PL($K$) diagram (Fig.~\ref{fig_PLK}) many of the C-rich Miras fall below
the anticipated PL relation, because circumstellar extinction is
sufficiently strong to affect their $K$ magnitudes, in some cases by more
than one magnitude.  This is illustrated by Nsengiyumva (2010 his fig.~3.9)
in a plot of the difference between $K$ observed and predicted from the PL
relation, as a function of $J-K$ colour.  

The O-rich Miras fall in approximately the same region as do the C-rich ones
at shorter periods, but above $\log P>2.7$ they are consistently brighter
than the PL relation.  These are the same stars that fall to the upper left
of the C stars in the colour-magnitude diagram (Fig.~\ref{fig_cm2}).  The
same is found for Miras in the LMC as discussed by Whitelock et al.  (2003),
who suggested that the high luminosity of these O-rich stars was a
consequence of hot bottom burning, which is expected for intermediate mass
AGB stars.  Feast (2009) has suggested that their position in the PL
relation is consistent with their being overtone pulsators, which may
eventually evolve into long period OH/IR stars.

In view of the fact that these O-rich Miras do not have significant
circumstellar reddening we can use the PL($K$) relation derived by Whitelock
et al.  (2008) to derive a distance.  For this we use only the four stars
with $P<400$ days, as longer period O-rich Miras are usually brighter than
the PL($K$) would predict.  We use the relation derived for Galactic stars
by Whitelock et al.  (2008): $$M_K=3.51[logP-2.38]-7.37,$$ which assumes the
same LMC distance as above.  Transforming the 4 $K$ magnitudes onto the SAAO
system as before gives a distance modulus for NGC\,6822 of $23.38\pm0.16$
mag.

Given the uncertainties, including the fact that the LMC  distance may vary
with the sample of LMC stars studied (e.g. Feast et al. 2012), the various
estimates of the NGC\,6822 distance modulus are not in conflict.

\section{Comparison with the Dwarf Spheroidals}

\begin{figure}
\begin{center}
\includegraphics[width=6cm]{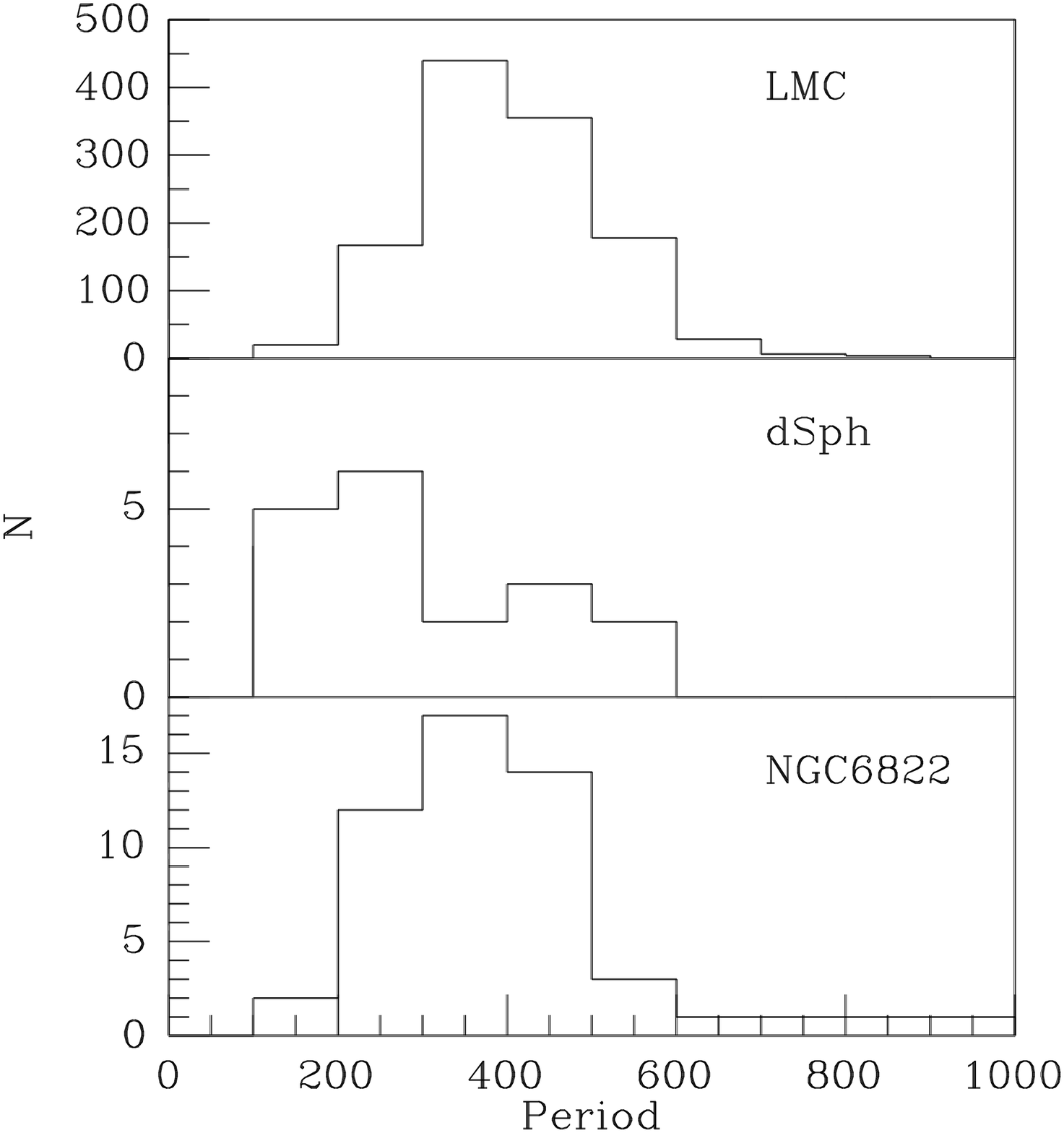}
\caption{Histogram of the periods of the C-rich Miras in 
NGC\,6822 (lower panel including N40419 P=193 days and BD~v24 P=670 days) and in the 
four dwarf spheroidals (central panel) and LMC (top panel from Soszy\'nski
et al. (2009)).}
\label{fig_hist}
\end{center}
\end{figure}

It is instructive to compare what we find here with the period distribution
of Miras in other Local Group galaxies and in particular with those in the
dwarf spheroidals that were surveyed in the same way.  Figure \ref{fig_hist}
shows a histogram of the periods for the presumed C-rich Mira variables and
compares them with those found in the dwarf spheroidals.  The following
dwarf spheroidals are involved (Mira periods, in days, given in parenthesis
after the names): Sculptor (189, 554, Menzies et al.  2011), Fornax(215,
258, 267, 280, 350, 400, 470, Whitelock et al.  2009), Phoenix (425 Menzies
et al.  2008), Leo~I (158, 180, 191, 252, 283, 336, 523, Menzies et al. 
2010) and Leo~II (183 unpublished).  We include the Phoenix dwarf galaxy
with this group, but note that it is generally classed as intermediate
between dwarf irregular and dwarf spheroidal (e.g. Battaglia et al.  2012).

Figure \ref{fig_hist} also shows the period distribution of probable C-rich
Miras in the LMC taken from Soszy\'nski et al.  (2009), which is based on
the OGLE~III catalogue.  Care is required in comparing frequency
distributions of AGB variables derived from surveys with different
sensitivities and in different wavelength regions (OGLE~III is a $V$ and $I$
survey), because the different selection effects will affect the colour- and
period-range of the variables found.  In particular longer wavelength
surveys tend to find larger numbers of redder and longer period variables.

The distribution of large amplitude carbon variables with period found for
NGC\,6822 is similar to that shown by carbon AGB variables in both the LMC
and SMC (see Soszy\'nski et al.  2011 fig.~3 for the SMC).  It is also clear
that there are very few short period carbon variables on the Wood sequence C
(the Mira sequence) of any amplitude in either Magellanic Cloud.  It is
notable that the distribution of LMC and SMC C-rich variables in both the
amplitude-period and the PL relations appear to be very similar though the
ratio of O-rich to C-rich variables differs greatly.

While the existence of Miras with periods in excess of 600 days in NGC\,6822
is quite striking, they only constitute 6 percent of the total number.  Thus
we would only expect to see one in all the Local Group dwarf spheroidals if
they were present in the same proportions.  However, our survey is probably not
complete for long-period variables, as our failure to identify BD~v21 (P=613
days) shows; it was too faint at $H$.

The smaller fraction of short period Miras, less than 300 days, in NGC\,6822
and the Magellanic Clouds compared to the dwarf spheroidals, is notable.  Of
course for short period Miras the magnitudes are fainter and the amplitudes
are lower (on average) than they are for the longer period stars, so they
are more difficult to find.  Nevertheless, we have followed the same
procedure here as we did in the dwarf spheroidals to identify variables, so
we should have found them if they were there, provided that they did not
have exceptional dust shells.  For example, BD~v16 (our N40590) was not
initially identified as a Mira by us, presumably because its amplitude was
slightly lower when we observed it than when they did.  It seems that the
variables in NGC\,6822 with periods less than 300 days have low amplitudes,
i.e., there are no stars like L7020 in Leo\,I which has $\Delta K_S =1.2$
mag and $P=191$ days.  However, L2077 in Leo\,I which has $\Delta K_S =1.2$
mag and $P=283$ days is much redder with mean magnitudes of $J=20.9$,
$H=19.0$ and $K_S=17.4$.  Given that our sensitivity is limited to stars
with $H<18.3$ mag this star would have been missed.
%Neither is there evidence for stars like this in the Galaxy or LMC (TRUE?).

Since the period decreases with increasing age for Miras, it seems probable
that the different period distributions are due to a larger proportion of an
older C-rich population in the dwarf spheroidals.  The evolutionary
status of this population is somewhat problematic, as discussed by Menzies
et al (2011).

%Why are the amplitudes of the short period (at least) Miras in the dwarf
%spheroidals greater than those of stars with similar period in NGC\,6822
%(lower T, higher mass-loss...)?

Although we have spectral types for only a few of the Miras in any of these
galaxies, the dwarf spheroidals do not have any long period M stars that we
know of. This of course is to be expected given what we know about the
metallicity and star formation history of the dwarf spheroidals.

\section{Conclusions}
 The large number of C-rich Miras now found in NGC\,6822 allows us to
demonstrate that the slope of the bolometric PL relation in that galaxy is,
within the errors, the same as that in the LMC.  The distance modulus found
from this relation is in satisfactory agreement with that found by other
methods and with that derived from the PL($K$) relation for the small number
of shorter period O-rich Miras.  Whilst there are problems with determining
bolometric magnitudes for C-rich AGB stars, these are not important for
distance scale studies provided a consistent method is employed for both
programme stars and calibrators.  The period distribution of high amplitude
carbon-rich AGB variables in NGC\,6822 is probably similar to that in the
two Magellanic Clouds but differs from that in Local Group dwarf
spheroidals, which contain a population of high amplitude, short period
C-rich variables.  Since these short period stars are believed to be old,
this indicates that the dwarf spheroidals contain an old population that is
capable of producing C-rich AGB variables which is absent or relatively rare
in both NGC\,6822 and the Magellanic Clouds.\\

\section*{Acknowledgements}

This publication makes extensive use of the various databases operated by
CDS, Strasbourg, France.  MWF, JWM and PAW gratefully acknowledge the
receipt of research grants from the National Research Foundation (NRF) of
South Africa and FN thanks the National Astrophysics and Space Science
Programme (NASSP) of South Africa for financial support.  We would also like
to thank Serge Demers for sending us data and preprints of his work with
Paolo Battinelli in advance of publication and the referee Jacco van Loon
for his comments.

\section*{Appendix}

\setcounter{figure}{0}
\begin{figure}
\includegraphics[width=8cm]{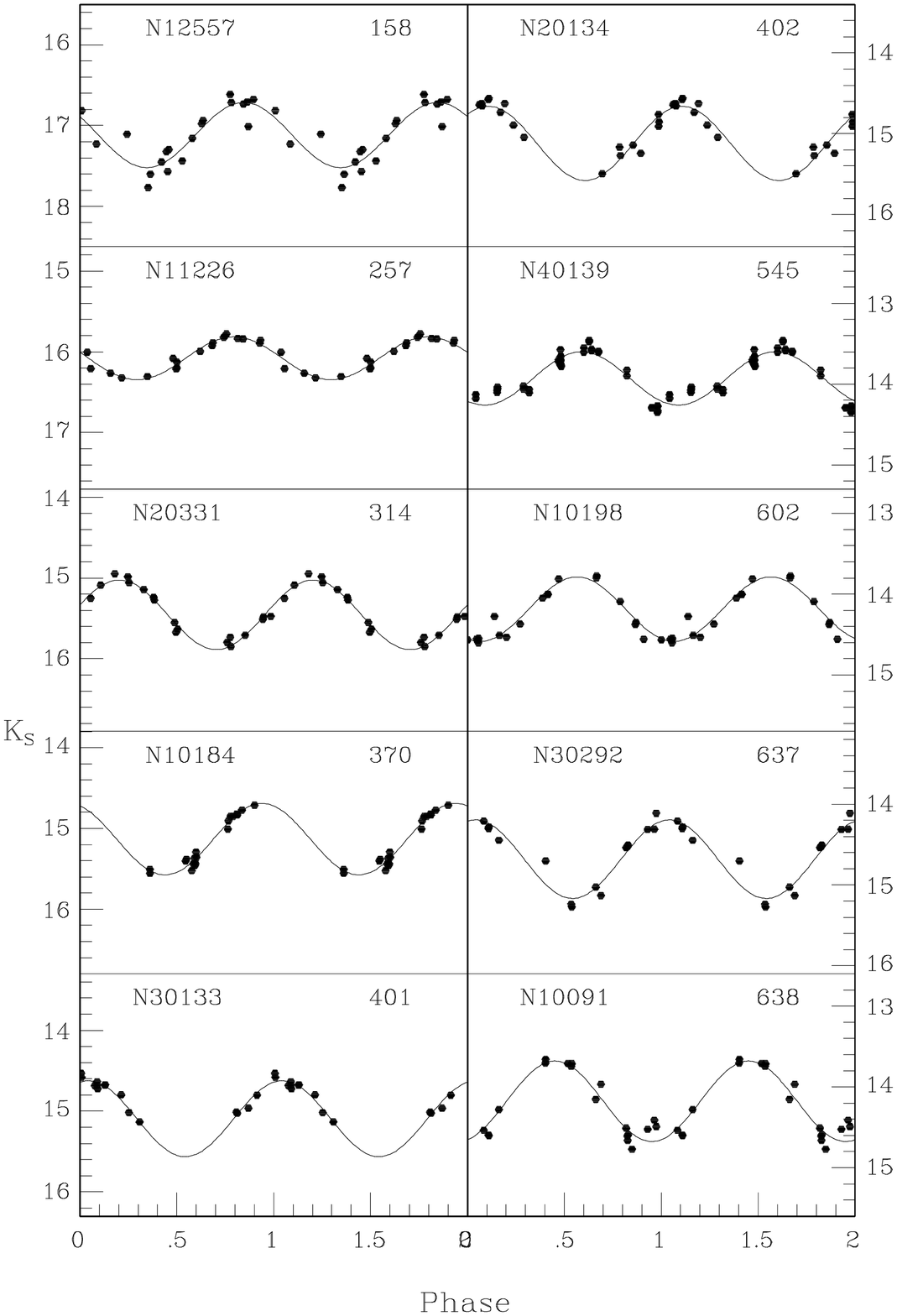}
\includegraphics[width=8cm]{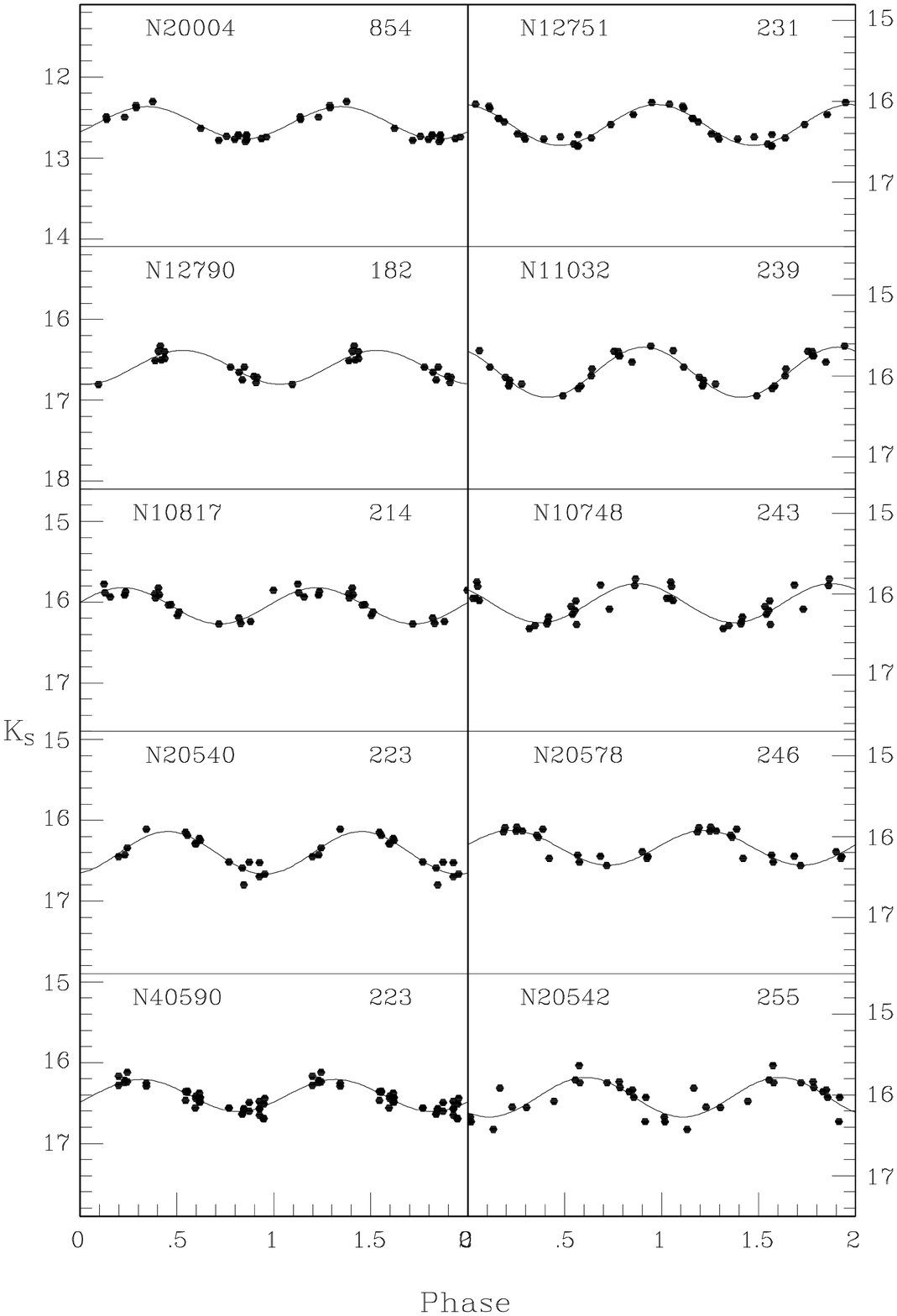}
\caption{$K_S$ light curves for the variables in
Table~\ref{tab_o}(a)(b), arbitrarily phased (zero at JD2450000); each point is plotted twice 
to emphasize the
variability. The best fitting first order curve is also illustrated, it is
from this that the mean magnitude and amplitude were determined.}
\label{fig_lc1}
\end{figure}
\setcounter{figure}{0}

\begin{figure}
\includegraphics[width=8cm]{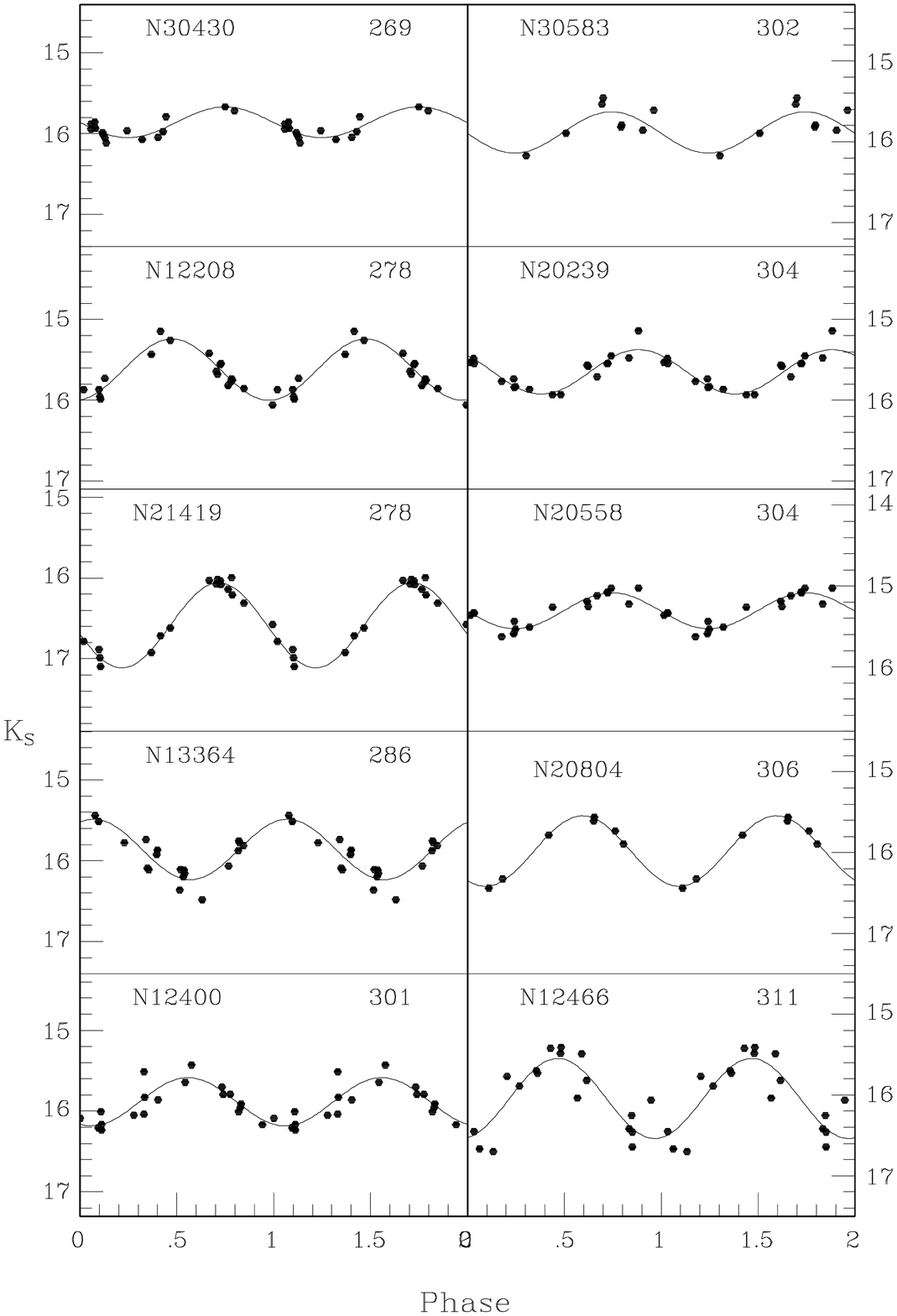}
\includegraphics[width=8cm]{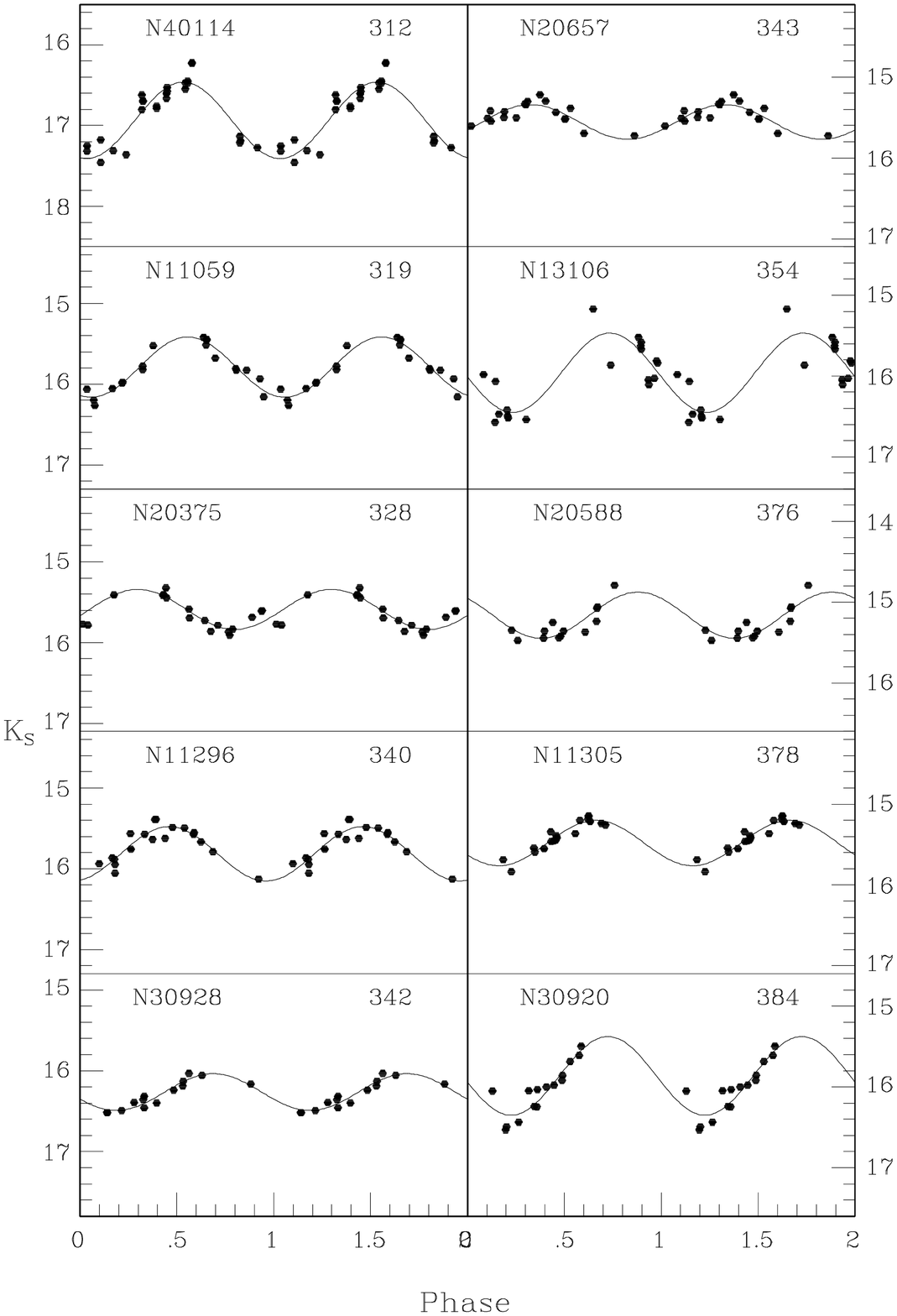}
\caption{continued}
%\label{fig_lc1}
\end{figure}

\setcounter{figure}{0}

\begin{figure}
\includegraphics[width=8cm]{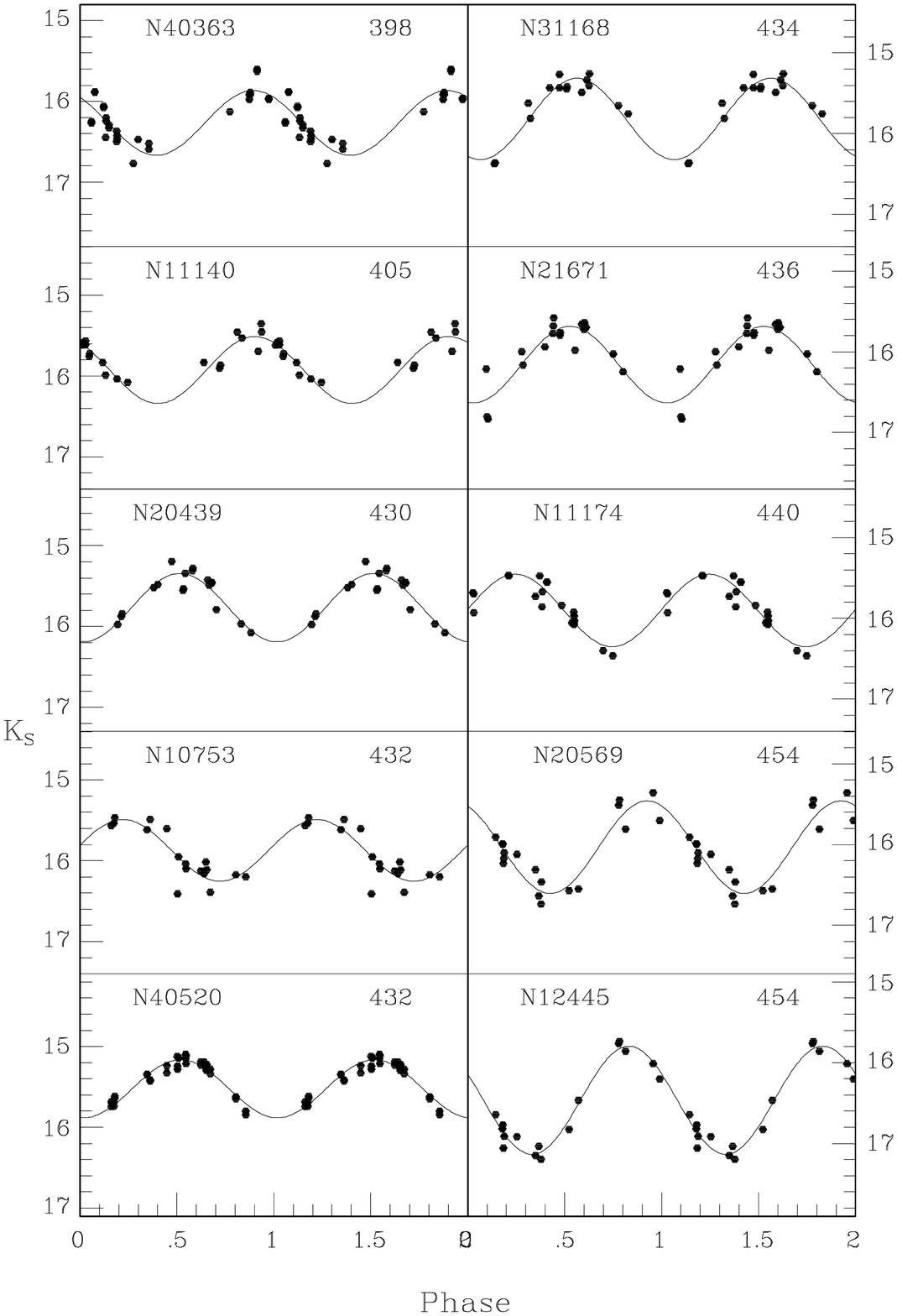}
\includegraphics[width=8cm]{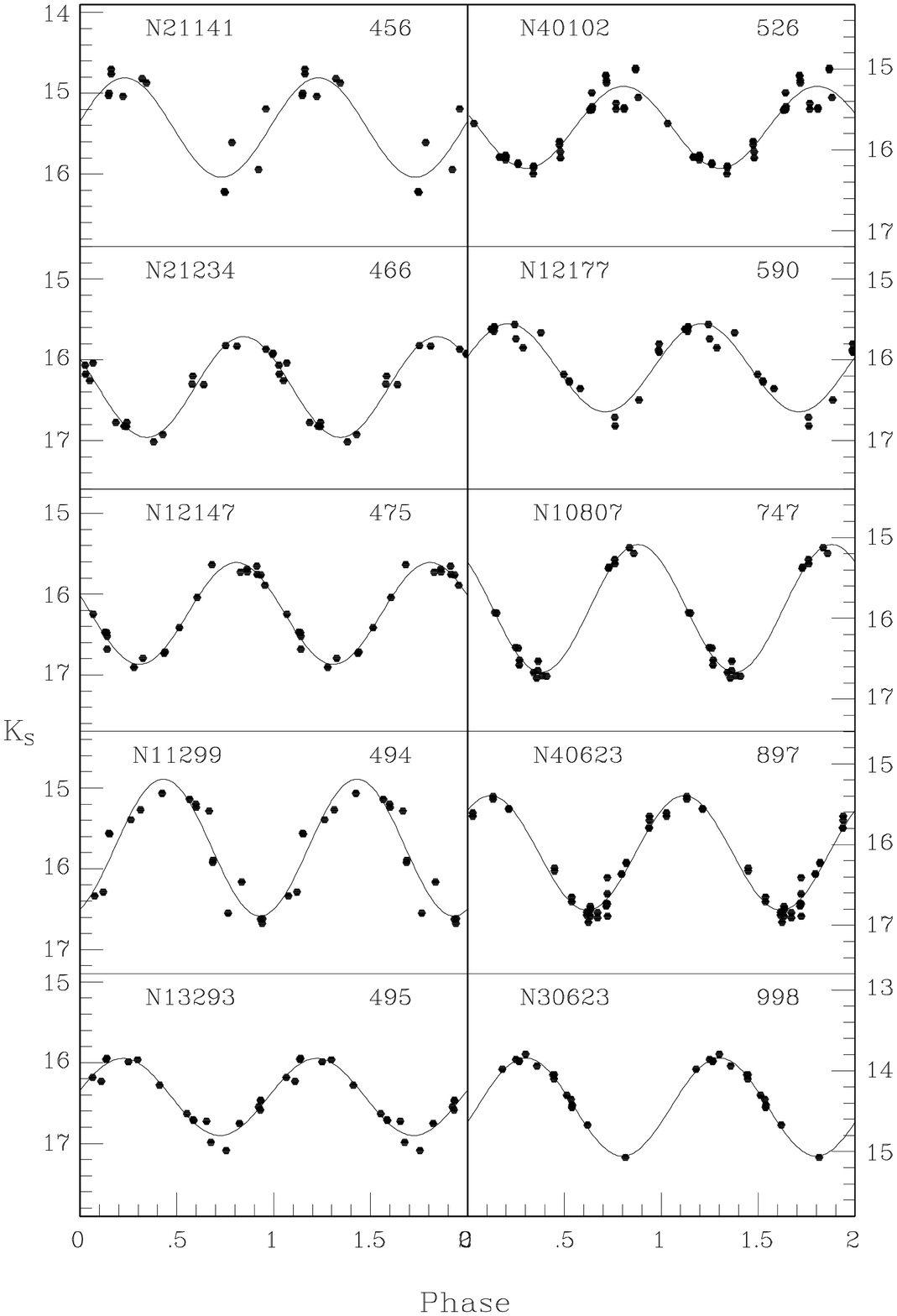}
\caption{continued}
\end{figure}

\begin{figure}
\includegraphics[width=8cm]{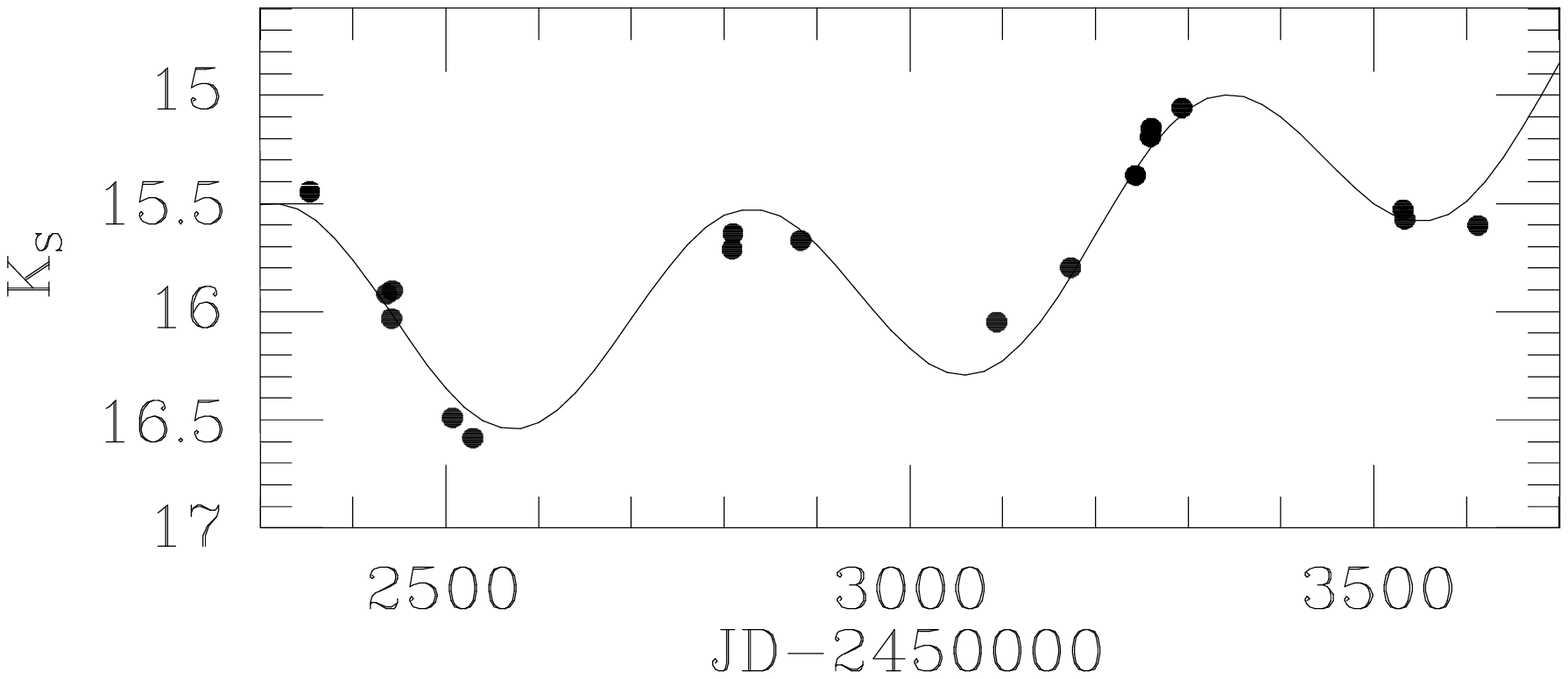}
\caption{$K_S$ light curve for N21029. The fitted curve is a combination of
two sinusoids, one with P=501 days, representing the pulsation, the other
with P=5000 days, representing a long-term trend. The latter is not a true
period simply an indication that there are secular or long period
changes in addition to the pulsation. }
\end{figure}

%\begin{figure*}
%\includegraphics[width=14cm]{fig_lc2.ps}
%\caption{$K_S$ light curves for the variables from
%Table~\ref{tab_o}(N20004 from (a) others from(b)). See the caption to 
%Fig.~\ref{fig_lc1} for details.}
%\label{fig_lc2}
%\end{figure*}

\end{document}